\documentclass[manuscript,screen]{acmart}
\AtBeginDocument{%
  }

\setcopyright{acmlicensed}
\copyrightyear{2018}
\acmYear{2018}
\acmDOI{XXXXXXX.XXXXXXX}
\acmConference[Conference acronym 'XX]{Make sure to enter the correct
  conference title from your rights confirmation email}{June 03--05,
  2018}{Woodstock, NY}
\acmISBN{978-1-4503-XXXX-X/2018/06}

\usepackage{caption}
\usepackage{subcaption}

\begin{document}

\title{Demographic Divides in Political Content Exposure on Facebook}

\author{S M Mehedi Zaman}
\email{sm.mehedi.zaman@rutgers.edu}
\affiliation{
    \institution{Rutgers University}
    \city{New Brunswick}
    \state{New Jersey}
    \country{USA}}

\author{Joao Couto}
\email{joaom.couto55@gmail.com}
\affiliation{%
  \institution{Federal University of Minas Gerais}
  \city{Belo Horizonte}
  \country{Brazil}
}

\author{Kiran Garimella}
\affiliation{%
  \institution{Rutgers University}
    \city{New Brunswick}
    \state{New Jersey}
    \country{USA}}
\email{kiran.garimella@rutgers.edu}

\renewcommand{\shortauthors}{Zaman et al.}

\begin{abstract}
  Despite Facebook's central role in American civic life, a clear, evidence-based understanding of users' long-term information environments has remained elusive, hindering assessments of the platform's societal impact. This study addresses that gap by analyzing a unique decade-long dataset, constructed by collecting the full list of public pages and groups followed by over 1,100 American users. This approach allows us to, for the first time, examine the potential information exposure of these users by analyzing hundreds of millions of posts from 2012 to 2023.

Our analysis reveals a complex information landscape. We find that political content constitutes a modest 18\% of a user's potential information diet, which is predominantly composed of lifestyle and entertainment topics. This aggregate view, however, masks a deeply stratified reality: we uncover significant and persistent disparities in the volume and ideological leaning of political content across age, gender, and racial lines. Furthermore, we quantify the porous boundaries between content categories, showing how political discourse frequently permeates non-political spaces. Leveraging the dataset's longitudinal nature, we also assess the impact of major platform interventions. We find, for instance, that Meta's 2018 ``Meaningful Social Interactions'' update dramatically increased the share of political content by contracting the visibility of non-political posts.

By providing a granular, decade-long map of potential information exposure, our study offers one of the first representative and longitudinal picture drawn from platform-independent data. Our findings underscore the critical need for researchers to measure exposure, not merely engagement, and to account for the significant volume of political content that circulates in non-political spaces. In an era of diminishing platform transparency, this work provides a vital empirical baseline for scholars and policymakers grappling with social media’s complex role in shaping modern civic life.
\end{abstract}



\received{20 February 2007}
\received[revised]{12 March 2009}
\received[accepted]{5 June 2009}

\maketitle

\section{Introduction}
Social media platforms have become central to how people learn, deliberate, and participate in civic life. Among them, Facebook remains the single largest online space for news and information exchange in the United States, used by a majority of adults across all demographic groups~\cite{pewresearchAmericansSocial}.

Yet, despite years of public debate about polarization, misinformation, and algorithmic influence, a basic empirical question has remained unresolved: \textit{What does the average American’s information environment on Facebook actually look like?} How much of it is political, how is it distributed across ideological lines, and how does it vary across demographic groups and over time?

Answering these questions is crucial because the composition of users’ information diets underpins broader debates about online polarization, the spread of misinformation, and the role of digital platforms in shaping public opinion. Without a clear, empirical baseline, public discussions often oscillate between two extremes: claims that social media are echo chambers of political outrage~\cite{garimella2018political}, and counterclaims that most users simply encounter mundane, apolitical content~\cite{barbera2020social}. A systematic measurement of political exposure is therefore a prerequisite for any serious understanding of social media’s political impact.

Investigating these patterns, however, is exceptionally difficult. The primary obstacle is the scarcity of granular data in what has been termed the ``post-API age'' of social media research~\cite{freelon2018computational}. Platforms tightly control the user-level data required for such analysis, forcing independent researchers to rely on methods with inherent limitations. Large-scale surveys, for instance, capture user demographics but are notoriously prone to self-reporting biases and fail to reflect the nuances of actual online behavior. Conversely, analyses of publicly available data via tools like CrowdTangle can reveal content trends but critically lack the demographic links needed to understand \textit{who} is being exposed to \textit{what}. These approaches provide either a fragmented or an aggregated view, failing to connect the supply of content to the diverse audiences that consume it.

This paper overcomes these challenges by constructing and analyzing a unique, decade-long dataset that maps the potential information exposure of a diverse panel of over 1,100 American users from 2012 to 2023. Our approach differs significantly from prior work by moving beyond self-reports and anonymized metrics. Using a privacy-preserving data donation methodology, we collected the full list of public pages and groups followed by our participants, retrieved over 193 million posts from these sources, and linked this content directly to verified user demographics. This allows us to measure, at an unprecedented scale and granularity, the composition and ideological character of the information environments that different demographic groups build for themselves.

Our analysis yields several key findings that challenge common narratives about social media. First, we reveal that explicitly political content comprises a relatively modest fraction of the average user's information diet, which is largely dominated by non-political topics like entertainment, hobbies, and community content. Second, we uncover profound and persistent demographic disparities in political content exposure: older users (55+) are exposed to significantly more political content compared to younger users (18-24), while we observe stark differences in the ideological leaning of content exposure across racial groups. These divides suggest that different communities inhabit fundamentally separate civic realities on the very same platform. 
Digging deeper into political content exposure, we find clear variation not only in how much political material users encounter but also in its overall leaning. While most of the political content is relatively centrist or neutral in tone, distinct partisan patterns emerge across demographic groups and over time.

Finally, we use our longitudinal data to study the impacts of  various popular platform interventions such as the 2018 ``Meaningful Social Interactions'' update~\cite{facebook2018} which down-ranked content from public Pages. We show that these interventions can have counterintuitive and powerful effects; the 2018 intervention, for example, dramatically increased the share of political content not by adding more, but by culling non-political content.
By providing one of the first population-level estimates of users' subscribed content on Facebook, our work offers a vital empirical foundation for a more nuanced understanding of the digital public sphere.

\section{Related Work}
\textbf{Studying Social Media in the Post-API Era}.
Independent research into the societal impact of social media is fundamentally constrained by the challenge of data access. The shift by major platforms to restrict or close their APIs has ushered in a ``post-API age," significantly complicating the ability of researchers to conduct large-scale, replicable, and transparent studies of online information environments~\cite{freelon2018computational}. This data scarcity forces a reliance on methods that often present an incomplete picture. Platform-provided tools like CrowdTangle, for example, have been invaluable for tracking the reach and spread of public content and have been used in thousands of academic studies~\cite{crowdtangleAboutCrowdTangle}. However, these tools were designed without the primary needs of academic researchers in mind; their key limitation is an inability to link content to the specific demographics of the users engaging with it, a critical gap this study directly addresses.

Data donation has emerged as a promising solution to this impasse~\cite{ohme2023digital}. By enabling users to voluntarily and consciously share their data for research, this approach facilitates the collection of high-quality, consent-based behavioral data with a well-defined sampling frame. This method carries immense ethical benefits, as it is rooted in an opt-in model with clear consent processes. However, it is not without its own challenges, including the logistical hurdles of recruiting a diverse user panel and the risk of non-cooperation from platforms that may hinder data collection efforts~\cite{keusch2023you}. Our study builds upon this growing body of work by implementing a robust data donation framework designed to construct a unique, longitudinal dataset that overcomes the core limitations of both platform-centric and survey-based research.

\textbf{The Composition of the American Information Diet}.
A central debate in media studies revolves around the composition of the average citizen's information diet and the role of political news within it. Our finding that political content constitutes approximately 18\% of users' potential exposure on Facebook provides a key empirical data point in this conversation. This figure is notably consistent with broader analyses of cross-media consumption, such as the work by \citet{allen2020evaluating}, which found that news accounts for 14\% of the average American's total media diet across various platforms, with television remaining a dominant source.

At the same time, this level of exposure on Facebook stands in contrast to other platforms. For example, research on Twitter has found that political content is far more prevalent, accounting for as much as one-third of all tweets posted by U.S. adults~\cite{pew2022twitterpolitics}. This comparison suggests that different platforms occupy distinct niches within the larger information ecosystem, and a one-size-fits-all assumption about social media usage can be misleading. Furthermore, our finding that the majority of content in users' subscribed ecosystems is non-political---dominated by lifestyle, entertainment, and religion---supports research arguing that most Americans' online media consumption is more moderate and less uniformly politicized than is often portrayed in popular discourse. This challenges the narrative of a hyper-politicized online environment and suggests that for many, Facebook's primary function remains social connection and personal interests.

\textbf{Demographic Divides and Incidental Political Exposure}.
While political content may represent a minority of the total content available, its distribution is far from uniform across the population. Our work confirms and extends a rich body of literature demonstrating that online information exposure is deeply stratified along demographic lines. This sorting of users into different information streams, where online behavior is predictable by age, gender, and ethnicity, is a key component of research on political polarization and the formation of ``echo chambers"~\cite{costello2016views,hobolt2024polarizing}. Our finding that older users are exposed to significantly more political content, for example, is consistent with studies that have identified age as a key predictor of sharing political news and being susceptible to online misinformation~\cite{guess2019less,sultan2024susceptibility}.

Furthermore, the ideological segregation we observe in users' subscribed pages builds upon foundational Facebook research which showed that while algorithms play a role, individual choice is a powerful driver of exposure to like-minded content~\cite{bakshy2015exposure}. 

However, exposure is not limited to explicitly political spaces. A significant body of research on incidental exposure has shown that users frequently encounter political information in seemingly non-political contexts, such as hobby groups, local community pages, or entertainment forums, often through content shared by their peers~\cite{weeks2017incidental}. This dynamic makes the boundaries between content genres porous, meaning that even users who do not actively seek out political news are still likely to encounter it. 

\textbf{The Impact of Algorithmic Curation on Information Ecosystems}
The information a user ultimately encounters on Facebook is shaped by a constant interplay between two powerful forces: user choice (the pages and groups they decide to follow) and algorithmic curation (the platform's system for ranking and displaying content in the News Feed)~\cite{bakshy2015exposure,flaxman2016filter}. Our study provides one of the first comprehensive maps of the former, offering a baseline ``universe of available content'' that users have explicitly opted into. This long-term, observational approach serves as a powerful complement to recent large-scale experiments that have sought to causally estimate the effects of the latter.

In 2023, a collaboration between Meta and academic researchers produced a series of studies designed to measure the algorithm's impact. For example, \citet{guess2023social} experimentally explored the behavioral and attitudinal effects of replacing the algorithmic feed with a chronological one, providing a direct test of the ranking system's influence. 
 
Recent work by \citet{talaga2025changes,fraxanet2025analyzing} provides direct evidence that Facebook's algorithmic changes between 2021 and 2024 led to a measurable decrease in the visibility of news content. Our decade-long dataset is uniquely positioned to identify and quantify the downstream effects of precisely these kinds of platform-driven trends.

\section{Data Collection}
\label{sec:data_collection}

To collect the data, we developed a Facebook application similar to popular platforms like Farmville~\cite{burroughs2014facebook}, designed specifically for the ethical collection of social media data through user consent. Users log in via our website using their Facebook credentials, selectively granting access to their interactions with groups and pages. This process leverages the Facebook Graph API, particularly focusing on endpoints such as user\_likes and groups\_access\_member\_info. Given the tightened security protocols post the Cambridge Analytica scandal~\cite{heawood2018pseudo}, our application underwent a rigorous manual approval process to justify each permission request, ensuring compliance with privacy standards. 

For each user, we get a list of Facebook groups and pages along with their basic demographics such as age, gender and ethnicity. 
We used CrowdTangle to obtain the posts shared in these pages/groups.
Apart from the demographics and the list of groups/pages for each user, we do not collect or store any other personally identifying information about the users donating the data. Our data collection received approval from the IRB at our university.

Even though there is a vast amount of public information on Facebook, including public pages and groups, accessing it presents significant challenges due to the exclusive control platforms have over data concerning the content exposed to users. Our data donation model allows us to bypass these platform restrictions and gain deeper insights into user behavior by allowing us access to get information on pages/groups from which users get their information.

We recruited users through an online survey panel company PureSpectrum. Purespectrum provides high quality panels and have been used in other academic studies~\cite{lazer2020covid}. The users donated their groups and took a 5 question survey asking them about demographics.
The entire process takes less than 3-5 minutes for the user. 
%
We obtained data from 1,115 users, which included a total of 251,220 pages and groups: approximately 210,000 pages and 42,000 groups. 
Our objective was to extract posts from these entities using CrowdTangle. However, due to CrowdTangle's limitation of 25,000 total entities per dashboard, we randomly sampled 10\% of both groups and pages, uploading them into separate lists on separate dashboards. Of these, around 17,000 of them have at least one post.\footnote{Note: we use the term Pages for simplicity and not Pages/Groups every time.}
A note on coverage by CrowdTangle: According to the FAQ page~\cite{crowdtangleWhatData}, any page that is not private is accessible via the CrowdTangle API. Thus, our dataset covers all public pages irrespective of their following. We miss a small fraction of pages (around 10\%) which are private.

The final dataset, comprises 193,383,571 posts published between January 2012 and July 2023.\footnote{CrowdTangle was shut down in August 2024. Due to the strict rate limits on CrowdTangle, we only present analysis with data we managed to collect up until that date.} Of those, 187,099,796 posts are tagged as English or Spanish according to CrowdTangle's `language\_code' field. It is noteworthy that 4,860,906 posts in the dataset have `und' (undefined) as their language code. The other remaining 1,422,869 posts are in a different language than English or Spanish. Within the `und' posts, there are some written in English; however, this study only considers English and Spanish posts as majority of the undefined language code posts were either emoji heavy or simply gibberish.



\begin{figure}[ht!]
    \centering
    \includegraphics[width=0.8\linewidth]{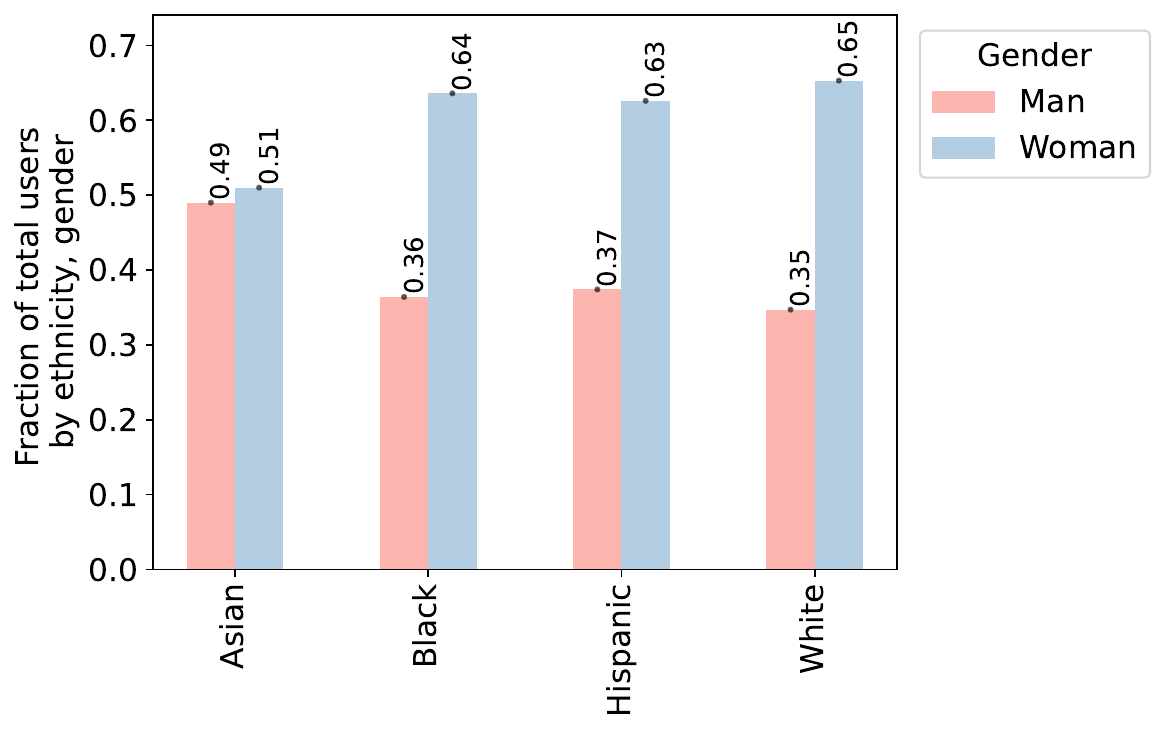}
    \caption{Fraction of users by ethnicity and gender.}
    \label{fig:frac_users_ethnicity_gender}
\vspace{-\baselineskip}
\end{figure}

\begin{figure}[ht!]
    \centering
    \includegraphics[width=0.8\linewidth]{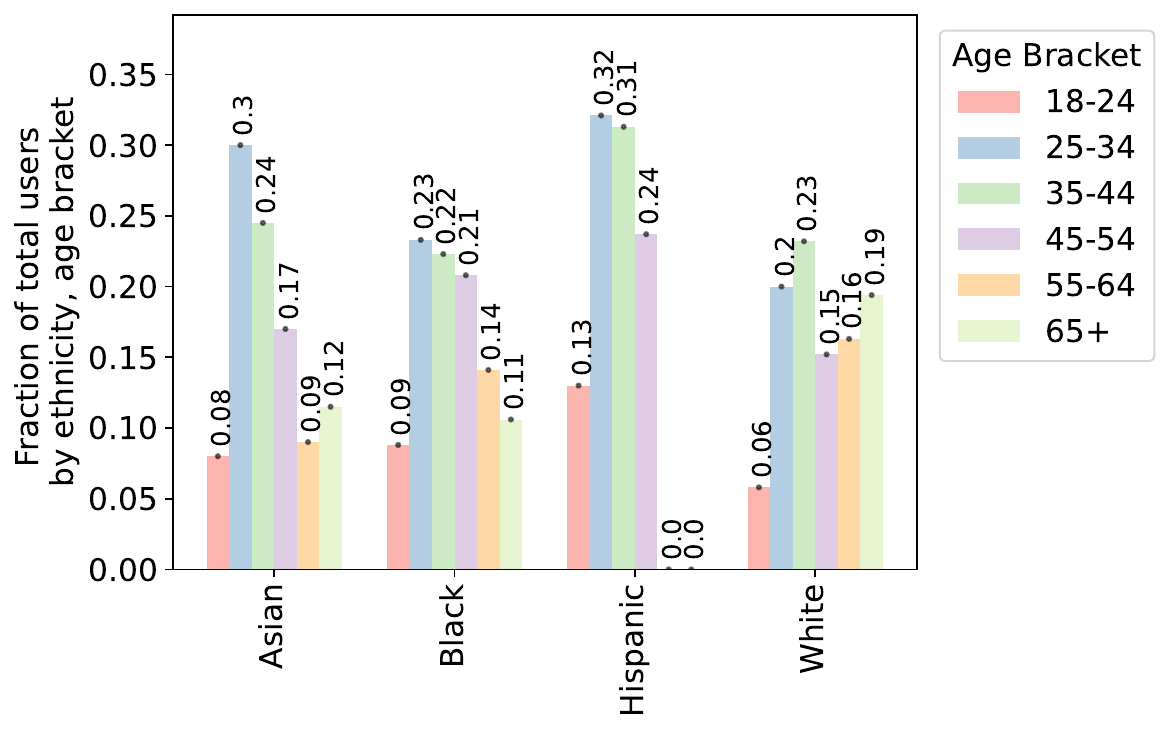}
    \caption{Fraction of users by ethnicity and age.}
    \label{fig:frac_users_ethnicity_age}
\vspace{-\baselineskip}
\end{figure}

The ethnic, gender and age distribution of our data is shown in Figures~\ref{fig:frac_users_ethnicity_gender},~\ref{fig:frac_users_ethnicity_age}.
We can see that we are oversampling women in most ethnicities and under sampling users in the 18-24 age category.
This is due to the convenient nature of our sampling. To compensate for the lack of representative data, we use reweighting techniques in our estimates in Section~\ref{sec:reweighting}.

\begin{figure}
    \centering
    \includegraphics[width=0.85\linewidth]{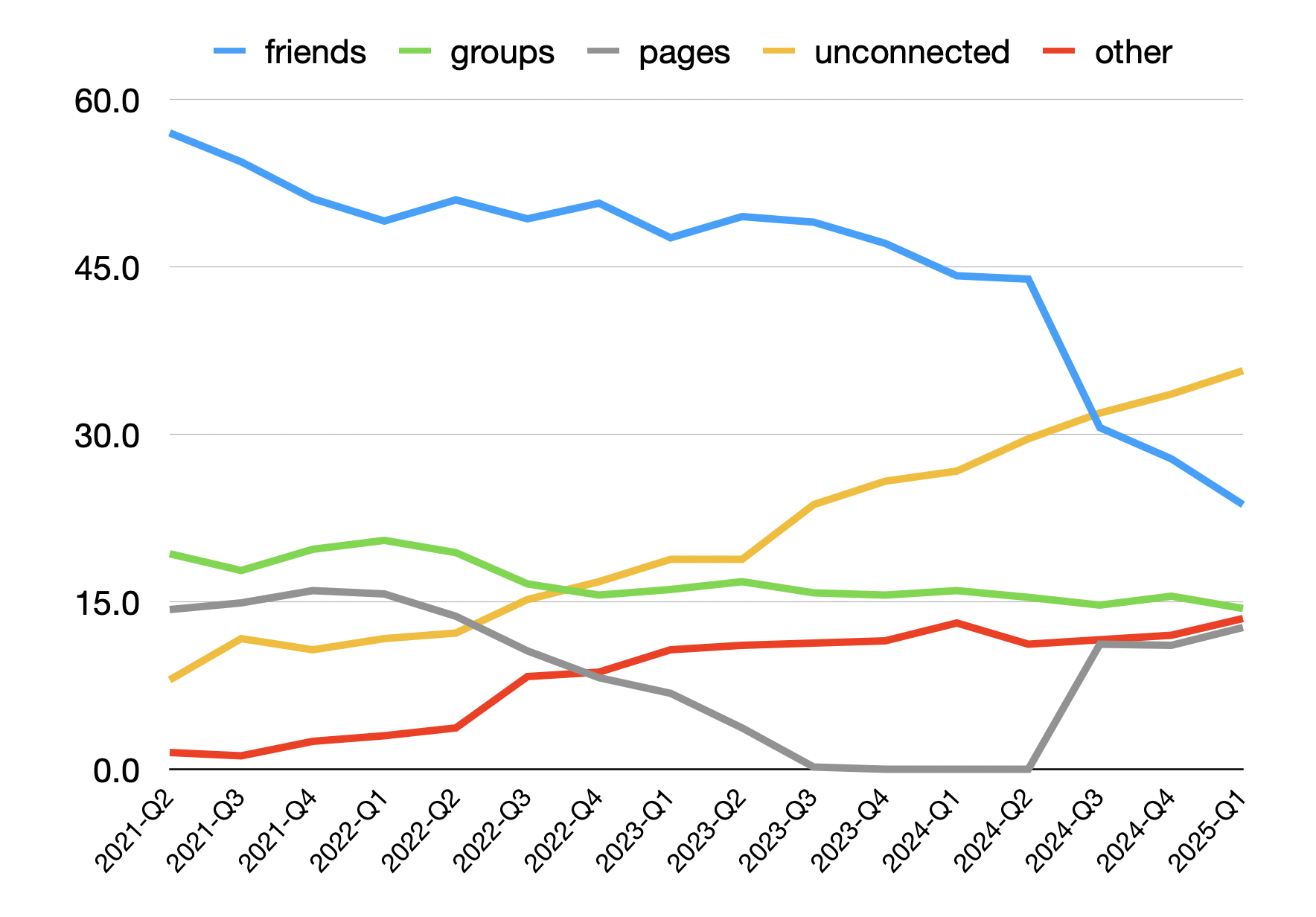}
    \caption{Fraction of content from various sources shown on user's feeds on Facebook.}
    \label{fig:facebook_frac_pages_groups}
\vspace{-\baselineskip}
\end{figure}

\textbf{A note on our sample}. Our dataset provides a robust measure of potential exposure, capturing the full universe of content produced by the public pages and groups that users have explicitly chosen to follow. While distinct from the final, algorithmically-curated content in a user's feed, this measure of user-selected information is a substantial component of their online experience. Indeed, Meta's own Widely Viewed Content Reports confirm that during the 2021-2022 period, content from followed sources consistently constituted 30-40\% of feed views—a figure that was likely even higher in the preceding years (Figure~\ref{fig:facebook_frac_pages_groups})~\cite{MetaTransparencyCenter2024}. This validates our approach, which aligns with established methods for inferring social media behavior in the absence of direct feed data~\cite{eady2023exposure}.

Methodologically, our sample offers three unique analytical lenses: it encodes user interest and preference through follows (each user’s chosen information ecosystem), it maps the full set of available posts from those sources so we can compare supply across demographic groups and over time (for example, whether the content available to White users is more right-leaning), and it provides a ten-year longitudinal record that lets us track how exposure evolves around major events and platform interventions. Combined with population reweighting, these data yield representative estimates of potential exposure while making clear what we do and do not observe: we study the supply from subscribed public sources rather than the exact ranked feed, which makes our estimates conservative but still highly informative for understanding the political information environment.

\textbf{A note on reproducibility}: The code for the data donation tool, the dataset containing the demographics, names/IDs of pages/groups, keywords used for identifying political/non-political pages, along with fine-tuned models for detecting political content will be released upon paper acceptance.
Unfortunately CrowdTangle's terms of service prohibit the release of the full raw dataset we collected,\footnote{\url{https://help.crowdtangle.com/en/articles/3192685-citing-crowdtangle-data}} but we hope the IDs and other artefacts might be valuable and can be used on Facebook's new open research and transparency tools for academics: the Meta Content Library~\cite{10.1145/3720533.3756893}.

\section{Data Processing}

In this section, we describe the various pre-processing steps taken before our analysis. We first annotated a subset of the groups for high precision political content.

A significant limitation of much of the existing research on online content exposure is its reliance on high-level, categorical labels. Studies often classify an entire page, group, or news domain as ``political" or a source of ``misinformation" and then assume all content from that source shares the same characteristic~\cite{guess2019less,talaga2025changes}. This approach, while useful for broad-scale analysis, is an oversimplification that misses a crucial dynamic of social media: the porous boundaries between content types. It cannot account for political discussions emerging in apolitical spaces or for non-political content being shared within partisan communities.

To overcome this limitation, our study employs a more granular, dual-level classification approach. First, we performed a page-level analysis where a large language model classified the titles of over 17,000 pages and groups into broad categories such as Political or Non-political. Second, we used a separate, fine-tuned classifier to analyze the text of every individual post within those pages. This dual approach allows us to precisely quantify the extent of content ``crosstalk", measuring how frequently users are exposed to political material in non-political environments, and vice versa.

\subsection{Labeling of Political and Non-Political Pages}
\label{page_level_politics_model}

To systematically study the prevalence of political content across Facebook pages, we classified all 17,053 pages into four categories: \textit{political}, \textit{non-political}, \textit{news}, and \textit{other}. We separated \textit{news} from \textit{political} because news outlets often publish political content but are not themselves political actors or advocates. This distinction helps separate political engagement from informational coverage.

We provided the page titles to GPT-5~\cite{openai2025gpt5} for classifying the pages into the four categories.

We verified 100 random samples from each group, finding about 95\% accuracy for \textit{political} and \textit{non-political} pages, 98\% for \textit{news}, and 82\% for \textit{other}. To improve reliability, we manually reviewed the remaining 3,206 \textit{other} pages to ensure proper classification. After all quality checks and revisions, our final dataset contained 458 \textit{political}, 12,896 \textit{non-political}, 1,051 \textit{news}, and 2,648 \textit{other} pages.
The \textit{other} category includes pages that were ambiguous and could not be reliably classified based on their names alone.
Next, we proceeded to label individual posts as political or non-political.

\subsection{Detection of Political Content}
\label{political_model}
We built our political or not classifier by utilizing the LLaMa 3.3 70B Instruct model \cite{meta_llama330_70BInstruct_2024}. Past works have demonstrated that LLMs are able to produce textual annotations that closely match human annotators \cite{ding-etal-2023-gpt}, particularly through instruction tuning. Hence, we utilize this decently large model to classify if a textual post is political or not. To make sure that the model will do a good job of labeling correctly, we first sampled 100 posts from each page and from those, we curated 250 political posts and 250 non-political posts ourselves. Then we tested these 500 samples with the model and it had a very good accuracy score of 0.97 with an F1-score of 0.95. Hence, we proceeded to label the entirety of our dataset (English and Spanish posts) of around 187M posts, which gave us about 62M political posts and 125M non-political posts. 

Despite the fact that text posts constitute only a minor fraction of the overall content (as illustrated in Figure~\ref{fig:content_types_age_groups}), nearly all types of posts on Facebook are accompanied by some form of text. Consequently, the coverage of our political content classifier remains highly effective, covering posts across various formats. This includes posts where the primary content may be an image or video; the accompanying text captions are utilized for classification purposes, ensuring broad applicability of our analysis tools regardless of the post type.

\subsection{Estimating left-right ideology}
\label{sec:llama}
We used the same LLaMa 3.3 70B instruct model to estimate the positioning of political texts in our dataset along the left-right ideological spectrum. Our approach involves directly asking the LLM for the positioning of individual posts whose textual content was previously identified as political.

To assess the reliability of our political labeling approach with LLaMa, we evaluated the model on two established datasets of political statements. Using the TwinViews-13k dataset~\cite{fulay-etal-2024-relationship}, which pairs left- and right-leaning statements across major political topics, the model achieved 98\% accuracy in distinguishing ideological leanings. We also tested it on statements derived from Pew Research’s ``Political Values of Harris and Trump Supporters'' dataset~\cite{pew_harris_trump_supporters_2024}, achieving 84.5\% accuracy. These evaluations confirm that our LLaMa-based labeling provides consistent and reliable political classifications across different contexts.

\subsection{Obtaining population level estimates}
\label{sec:reweighting}

Because our dataset was based on a convenience sample with demographic quotas, we applied reweighting~\cite{sarig2023balance} to produce population-level estimates. We generated propensity scores using demographic variables such as age, gender, and ethnicity, and used these to adjust the sample distribution to better match the target population. This weighting procedure corrects for sampling bias and allows us to derive representative statistics despite non-random data collection. All results reported in the following sections are weighted to reflect population-level estimates.

\section{Results}
\label{sec:results}

We organize our results around four key dimensions of political content exposure on Facebook. First, we measure how much political content users encounter overall and within ostensibly non-political spaces, providing a baseline for understanding the broader information environment. Second, we examine political content exposure across demographic groups to identify systematic differences by age, gender, and ethnicity. Third, we analyze the ideological leaning of the political content users are exposed to, highlighting how these patterns mirror known partisan divides in the United States. Finally, we study long-term trends in political content production and evaluate how major platform interventions---such as ranking and feed adjustments---have shaped both the volume and composition of political exposure over time.

\subsection{Political content in non political spaces}
Our first analysis examines how much political content users encounter in spaces that are not explicitly political. Much of the prior literature~\cite{guess2019less,talaga2025changes} assumes that political exposure can be measured based on whether a page or domain itself is political. However, this assumption overlooks the possibility that substantial political content may circulate in otherwise non-political environments.

\begin{table}[H]
\small
  \centering
  \caption{Summary of page categories, post volumes, and political content shares. “(\%) Political” denotes the share of political posts within each category, while “\% Political Posts” shows each category’s contribution to all political posts in the dataset. Despite having only 13.6\% political posts, non-political pages account for 30.8\% of total political content due to their large number and activity.}
  \label{tab:page-summary}
  \setlength{\tabcolsep}{3.5pt}      
  \renewcommand{\arraystretch}{1.15} 
  \footnotesize                      

  \begin{tabular}{lcccccc}
    \hline
    \textbf{Category} &
    \shortstack[c]{\textbf{\#}\\\textbf{Pages}} &
    \shortstack[c]{\textbf{\# Posts}\\\textbf{(M)}} &
    \shortstack[c]{\textbf{Political}\\\textbf{Posts (M)}} &
    \shortstack[c]{\textbf{\% Total}\\\textbf{ Posts}} &
    \shortstack[c]{\textbf{\% }\\\textbf{Political}} & 
    \shortstack[c]{\textbf{\% Political}\\\textbf{Posts}}\\
    \hline
    Non-political & 12{,}896 & 119.7 & 16.3 & 64.0\% & \textbf{13.6\%} & 30.8\%\\
    Political     &    458   &  12.2 &  8.6 &  6.5\% & \textbf{70.5\%} & 16.3\%\\
    News          &  1{,}051 &  39.5 & 23.1 & 21.1\% & \textbf{58.6\%} & 43.7\%\\
    Other         &  2{,}648 &  15.7 &  4.9 &  8.4\% & \textbf{31.1\%} & 9.2\%\\
    \hline
    \label{tab:page-summary}
  \end{tabular}
\vspace{-001em}
\end{table}

Table~\ref{tab:page-summary} summarizes the distribution of pages, posts, and political content across four categories: \textit{political}, \textit{non-political}, \textit{news}, and \textit{other}. Political content fractions were computed using post-weighted measures, representing the proportion of political posts within each category. As expected, pages labeled \textit{political} are heavily dominated by political material, with 70.5\% of their posts classified as political. This pattern validates our page-level classification. Yet these pages contribute only 16.3\% of all political posts in the dataset.

The majority of political content arises from two other sources: \textit{news} and \textit{non-political} pages. News outlets, where 58.6\% of posts are political, produce 43.7\% of all political posts observed. Meanwhile, non-political pages show only 13.6\% political content individually, but their large number means they collectively account for 30.8\% of all political posts. This “long tail” of incidental political expression represents a major share of users’ potential political exposure.

Overall, over 80\% of political content originates from sources that are not explicitly political. This finding challenges common methodological assumptions and underscores that the political information environment on Facebook extends far beyond overtly partisan pages. Users are exposed to politics both intentionally, through following political pages, and incidentally, through news and apolitical communities. The latter form of exposure may be particularly influential, as it occurs in contexts where users may not anticipate or critically evaluate political messaging. Understanding this broader landscape is essential for accurately capturing how political discourse permeates everyday online spaces.

\subsection{Political content exposure}

\begin{figure}[ht]
\centering
\begin{subfigure}[b]{0.22\textwidth}
    \includegraphics[width=\textwidth]{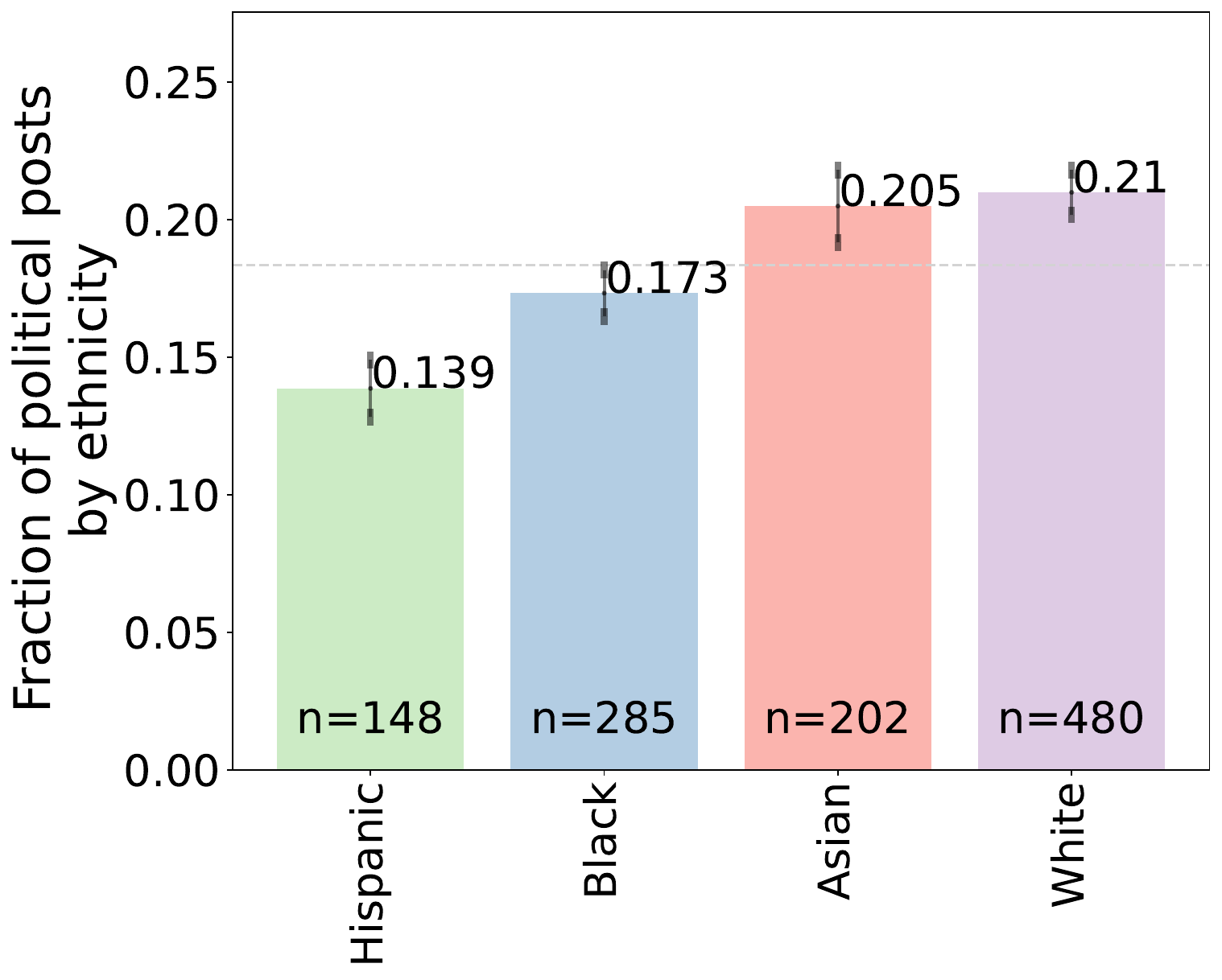}
    \caption{Ethnicity}
    \label{fig:political_ethnicity}
\end{subfigure}
\hfill
\begin{subfigure}[b]{0.22\textwidth}
    \includegraphics[width=\textwidth]{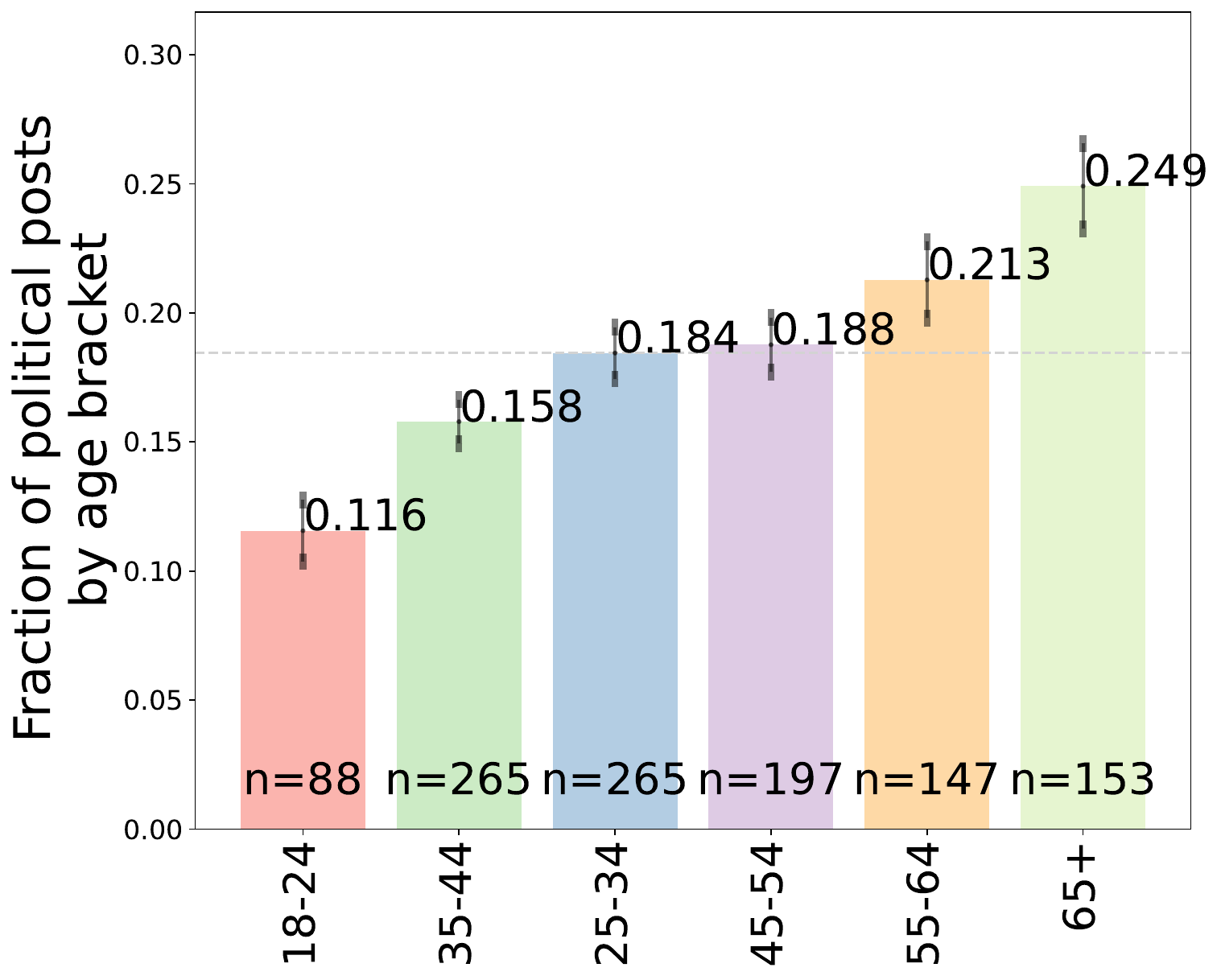}
    \caption{Age group}
    \label{fig:political_age}
\end{subfigure}
\hfill
\begin{subfigure}[b]{0.20\textwidth}
    \includegraphics[width=\textwidth]{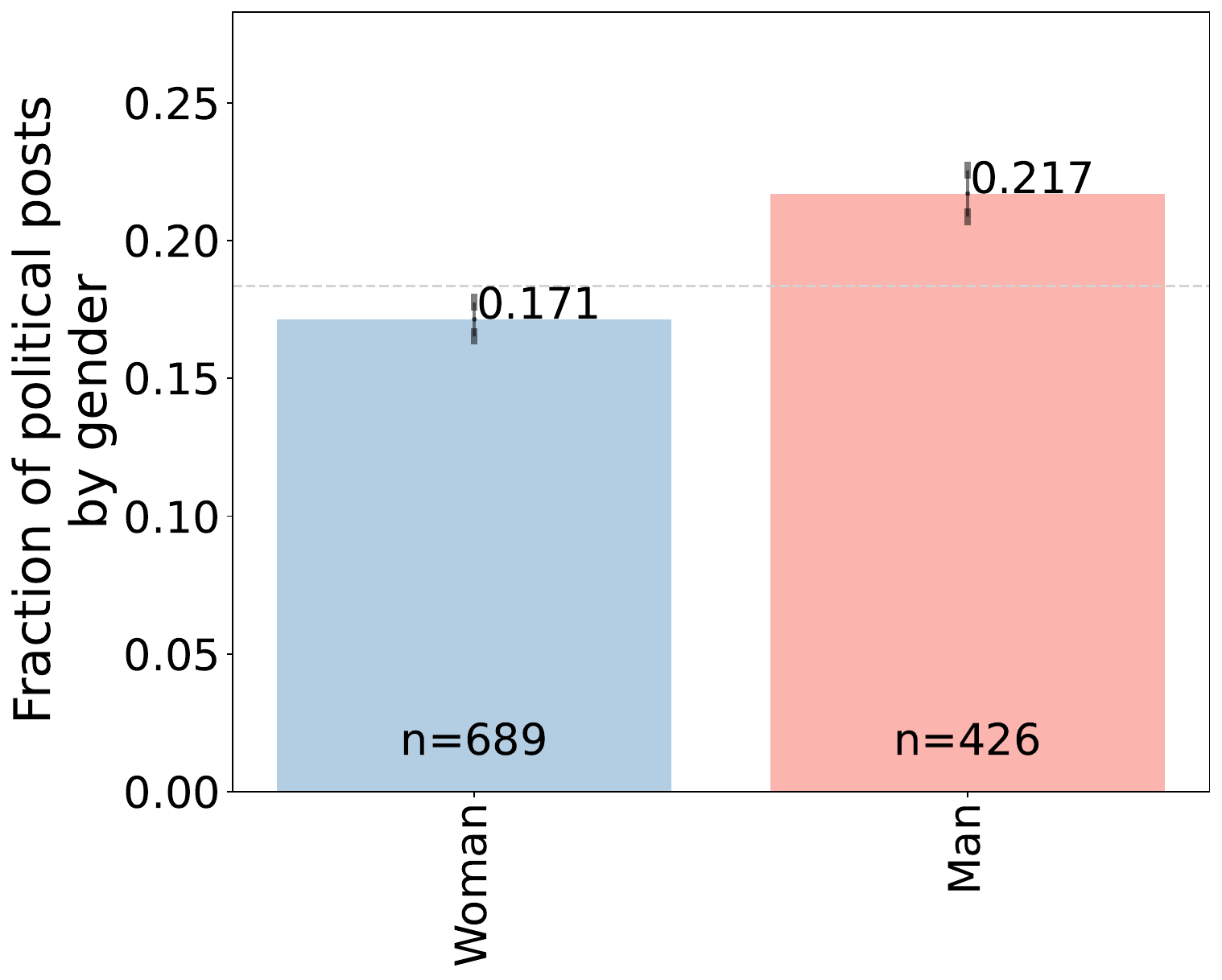}
    \caption{Gender}
    \label{fig:political_gender}
\end{subfigure}
\hfill
\begin{subfigure}[b]{0.26\textwidth}
    \includegraphics[width=\textwidth]{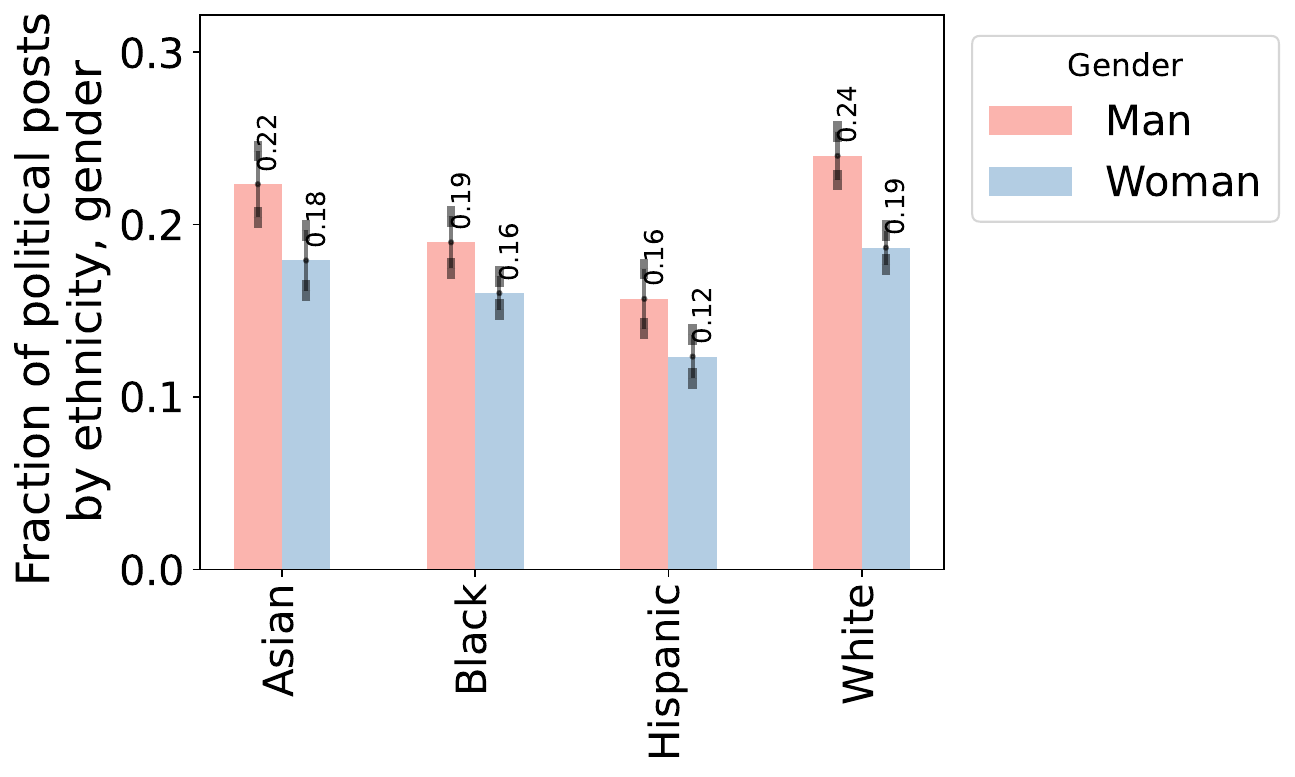}
    \caption{Ethnicity and gender}
    \label{fig:political_gender_ethnicity}
\end{subfigure}
\caption{Exposure to political content by: (a) ethnicity, (b) age group, (c) gender, (d) gender and ethnicity.}
\label{fig:political_content}
\vspace{-\baselineskip}
\end{figure}

This section examines the distribution and exposure to political content among various demographic groups. 
In each plot in figure~\ref{fig:political_content}, the dotted gray line indicates the overall population mean and the error bars indicate 95\% confidence intervals.
%
Overall, around 18\% of the potential content exposure is political content.
These findings align with broader media consumption trends identified in other studies on Twitter and television~\cite{pew2022twitterpolitics,allen2020evaluating}. 
Sections~\ref{subsec:topics_disproportionately} and~\ref{subsec:heterogeneity_content_consumption} (Appendix) detail heterogeneity in other topics exposed to various demographics.

Figure~\ref{fig:political_ethnicity} reveals a significant difference in political content exposure among ethnic groups. Hispanics engage with political content significantly less, at about 14\%, compared to Asians and Whites, each around 20\%. 
As depicted in Figure~\ref{fig:political_age}, age correlates strongly with potential exposure to political content. Younger individuals (18-24 years) show a markedly lower engagement rate at 12\%, in contrast to older users (65+ years), who engage at almost a double that rate, of 25\%. This finding is in line with previous research which showed that older users are significantly more active in consuming misinformation on Facebook~\cite{guess2019less}. The same pattern across age groups also holds for a `News' content --- as defined by Facebook's page/group categorization (See Figure~\ref{fig:news_consumption_age} in the Appendix).
Figure~\ref{fig:political_gender} indicates that men are potentially exposed to more political content than women. This finding is consistent across ethnicities (See Figure~\ref{fig:political_gender_ethnicity}).
%
%
Figures~\ref{fig:10y_political_age},~\ref{fig:10y_political_eth} show the timeseries of trends spanning 10 years for age and ethnicity respectively. 
We can clearly see that the trends and differences across various groups (e.g. older users exposed to more political content, Hispanics exposed to the lowest political data, etc.) also persist across the 10 years.

This study provides the first detailed estimates of how much political content users are exposed to on Facebook. It shows that most of the content users potentially see is not political. These results are important because they challenge the common view of Facebook as a space dominated by political debate. The finding that political posts make up only about 18\% of the content in an average user’s feed is especially noteworthy.

\subsection{Political leaning analysis}

\begin{figure}[ht]
\centering
\begin{subfigure}[b]{0.23\textwidth}
    \includegraphics[width=\textwidth]{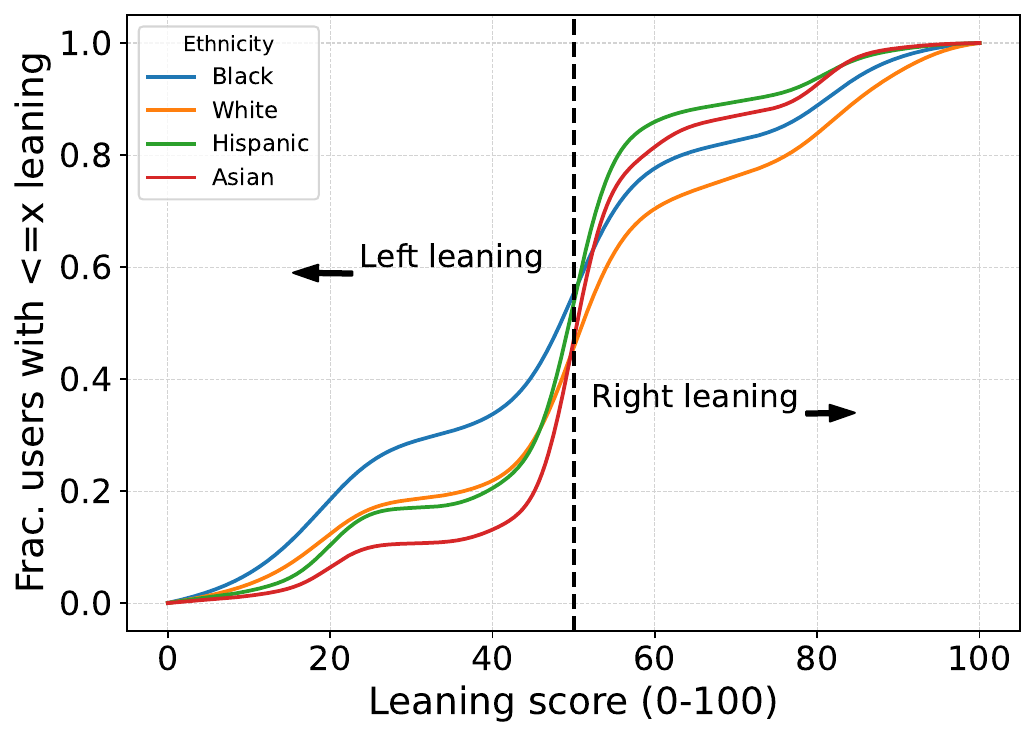}
    \caption{Ethnicity}
    \label{fig:political_ethnicity_leaning}
\end{subfigure}
\hfill
\begin{subfigure}[b]{0.23\textwidth}
    \includegraphics[width=\textwidth]{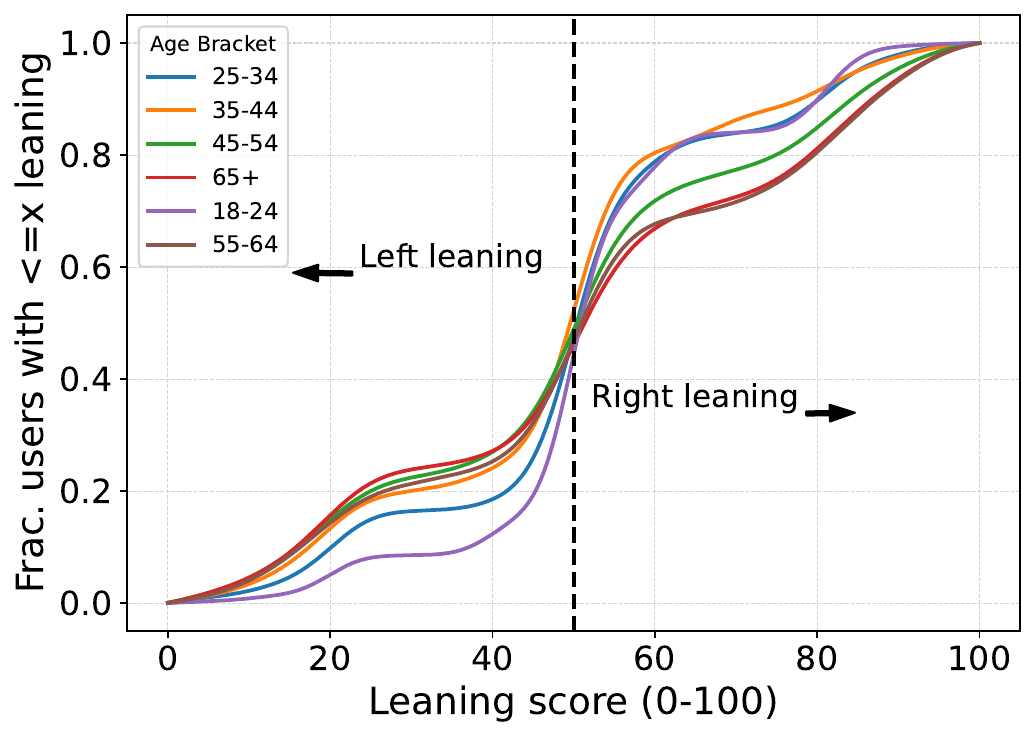}
    \caption{Age group}
    \label{fig:political_age_leaning}
\end{subfigure}
\hfill
\begin{subfigure}[b]{0.26\textwidth}
    \includegraphics[width=0.9\textwidth]{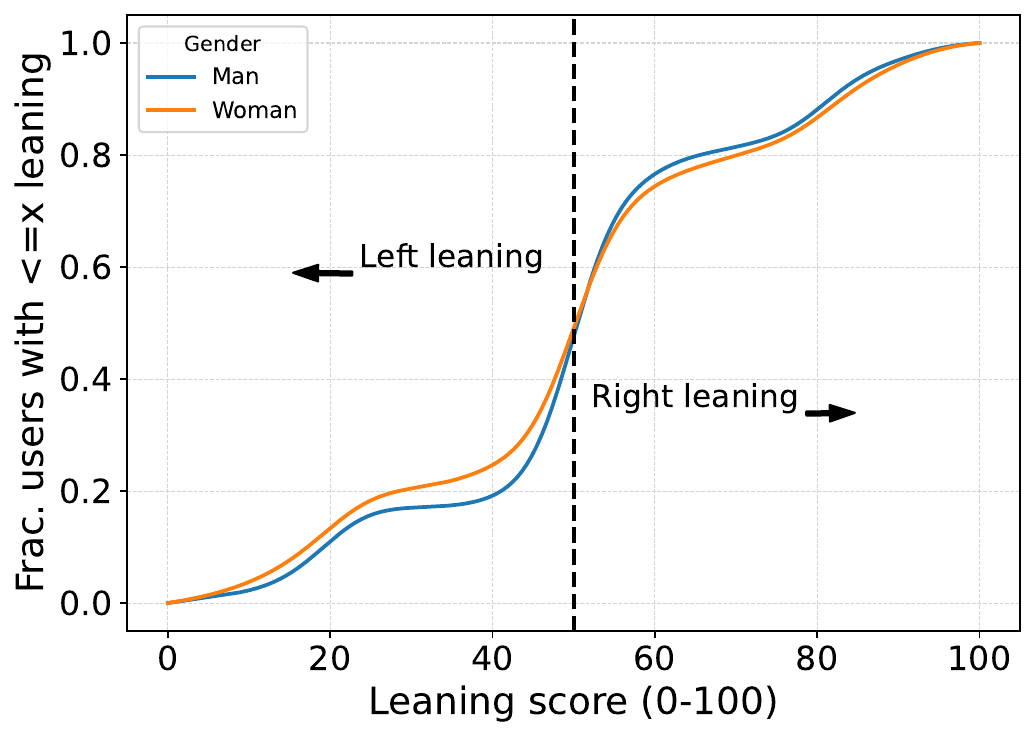}
    \caption{Gender}
    \label{fig:political_gender_leaning}
\end{subfigure}
\caption{Partisan content exposure leaning by: (a) ethnicity, (b) age group, (c) gender}
\label{fig:left_right}
\vspace{-\baselineskip}
\end{figure}

We next examine the political leaning of the content that different demographic groups are exposed to, not just the volume of political material. Following Section~\ref{sec:llama}, each political post receives a score in $[0,100]$ where lower values are left-leaning and higher values are right-leaning. We plot cumulative distribution functions (CDFs) of these scores by ethnicity, age and gender (Figure~\ref{fig:left_right}). For ease of interpretation, the CDF at a given $x$ shows the share of exposure at or below that leaning threshold. All estimates use the weighting procedure described in section ~\ref{sec:reweighting}.

A general pattern is that most exposure lies near the center: a large share of political posts across groups falls between 40 and 60, with many points clustered around 50. This aligns with our earlier finding that news and non-political pages account for most political posts in the dataset (Table~\ref{tab:page-summary}). Much of what users see is straight political news or informational coverage, which tends to be less extreme than content from overtly political pages.

By ethnicity, Figure~\ref{fig:political_ethnicity_leaning} shows clear but modest differences in partisan content exposure. Black respondents have more mass left of 50 indicating a majority democrat support for Black respondents. We see the opposite for White respondents, who are shifted right relative to other groups. These patterns are consistent with well-documented group differences in U.S. partisanship~\cite{pew2024changing}. 
By age, Figure~\ref{fig:political_age_leaning} displays a familiar gradient: 18–24 year-olds are the most left-leaning on average, 25–44 remain left of center but closer to neutral, 45–54 are near even, and 55–64 and 65+ shift right. By gender, Figure~\ref{fig:political_gender_leaning} shows small differences: women are slightly more left-leaning on average, but the gap is much smaller than for ethnicity or age.

To get clear trends in absolute percentages of potential exposure, we also computed binary exposure shares of explicitly partisan content with leaning score $<$40 as left leaning and $>$60 as right leaning. Figures~\ref{fig:left_right_overall_ethnicity},~\ref{fig:left_right_overall_age}, and~\ref{fig:left_right_overall_gender} (Appendix) show these trends. There is a clear difference in Black participants exposed to more left leaning content, younger people and men exposed to more right leaning content. These trends are also consistent across time with a clear decline in the 18-24 share of left leaning content since 2021 (Figure~\ref{fig:left_right_overall_age_time}, Appendix).

\subsection{Algorithmic impacts on content exposure}

We next study how major internal and external interventions relate to political content exposure on Facebook. We focus on three large platform changes that fall within our observation window: the June 2016 “Friends and Family” ranking change~\cite{FB_friendsfamily_2016}, the January 2018 “Meaningful Social Interactions” update~\cite{facebook2018}, and the January 2021 ``News and Civic Content Deprioritization"~\cite{nytimesFacebookDials2021}. 
These interventions adjust News Feed ranking toward personal and conversational content and away from publisher and civic posts. The 2016 update increases content from friends and family. The 2018 update favors posts that spark meaningful back-and-forth conversations. The 2021 update explicitly reduces the visibility of political and civic news.

We are interested in these updates because they change what surfaces in the feed and may shift both the level and the sources of political exposure and production. In particular, they can lower supply from news pages or raise the visibility of conversational posts from non-political pages, which may change incidental political exposure. Prior work has mainly tracked engagement outcomes such as likes and shares~\cite{talaga2025changes,fraxanet2025analyzing}. Our results complement that work by focusing on political content exposure.

\textbf{Methods}. We estimate the causal effect of each intervention on exposure to political content using a Bayesian structural time-series model (BSTS)~\cite{brodersen2015inferring}. The model learns the behavior of the outcome in the pre period and then predicts the counterfactual path that would have occurred in the post period if the intervention had not happened. In contrast to classical difference-in-differences, BSTS provides a full posterior for the effect over time, allows time-varying trends and seasonality, and incorporates predictive covariates as a synthetic control inside the state-space model~\cite{brodersen2015inferring}.

We study two outcomes at time $t$: the weighted fraction of political posts in the feed, $y_t^{\text{share}}$, and the weighted volume of political posts, $y_t^{\text{count}}$. For each intervention, we fit on the pre period and evaluate effects over post windows of 1, 2, 3, and 4 years.
We use a local-trend plus seasonal BSTS with dynamic regression terms:
$y_t$=$\mu_t$ + $\gamma_t$ + $x_t^\tau\beta_t$ + $\epsilon_t$,
where $\mu_t$ is a local level with slope, $\gamma_t$ captures weekly and yearly seasonality, $\mathbf{x}_t$ are covariates that are predictive but not affected by the intervention, and $\varepsilon_t \sim \mathcal{N}(0,\sigma^2)$. We include covariates such as calendar effects, overall activity levels, and stable composition controls. Spike-and-slab priors select useful covariates while shrinking the rest. Parameters are estimated by MCMC, which yields posterior draws for all coefficients.
After fitting the model to the pre period, we generate the posterior predictive distribution for $y_t$ in the post period conditional on covariates, as if no intervention occurred. This gives $y^{\text{cf}}_t$ with credible intervals. The pointwise effect is $\Delta_t$ = $y_t^{obs}$ - $y_t^{cf}$.
%
To simplify interpretation, we only report the relative effect, which is computed as the ratio of the cumulative absolute difference between the observed outcome and the model’s posterior counterfactual prediction, divided by the cumulative counterfactual prediction, expressed as a percentage.

\begin{figure}
    \centering
    \includegraphics[width=0.7\linewidth]{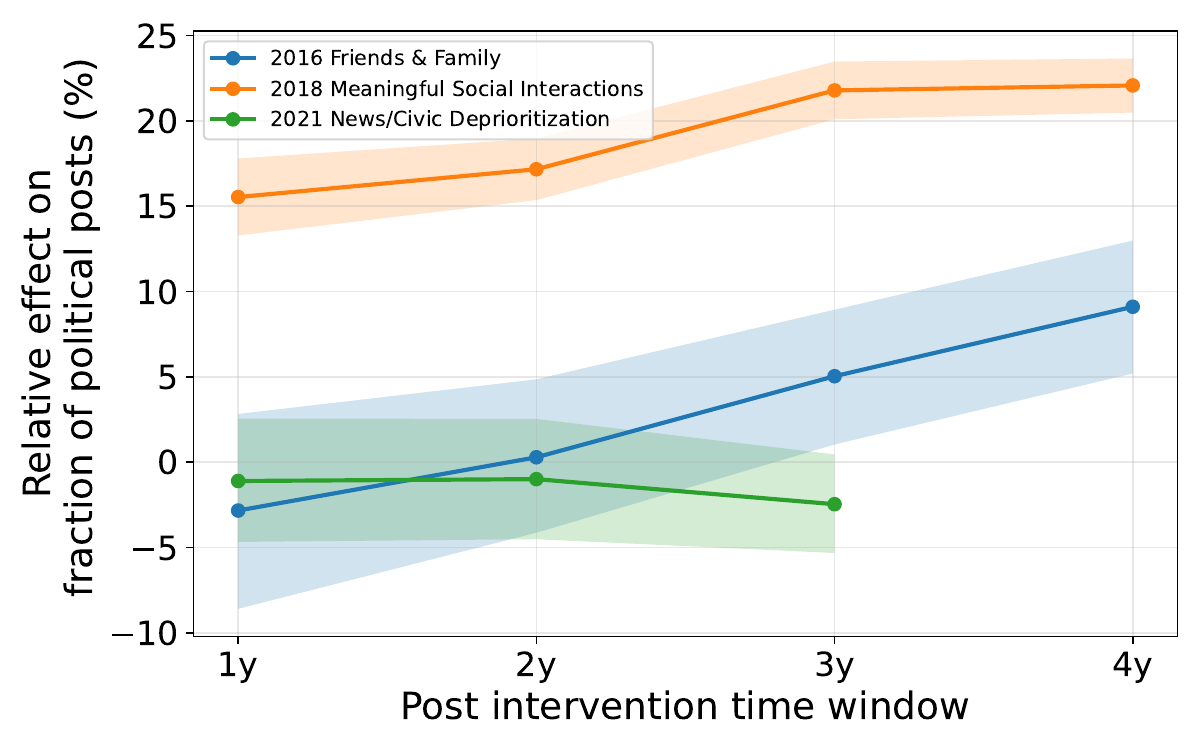}
    \caption{Relative effect of the three interventions on the \textbf{fraction} of political posts. Lines show posterior means and shaded areas show 95\% confidence intervals. Effects are relative percent differences from the BSTS counterfactual.}
\label{fig:interventions_combined_political_fraction}
\vspace{-\baselineskip}
\end{figure}

\begin{figure}
    \centering
    \includegraphics[width=0.7\linewidth]{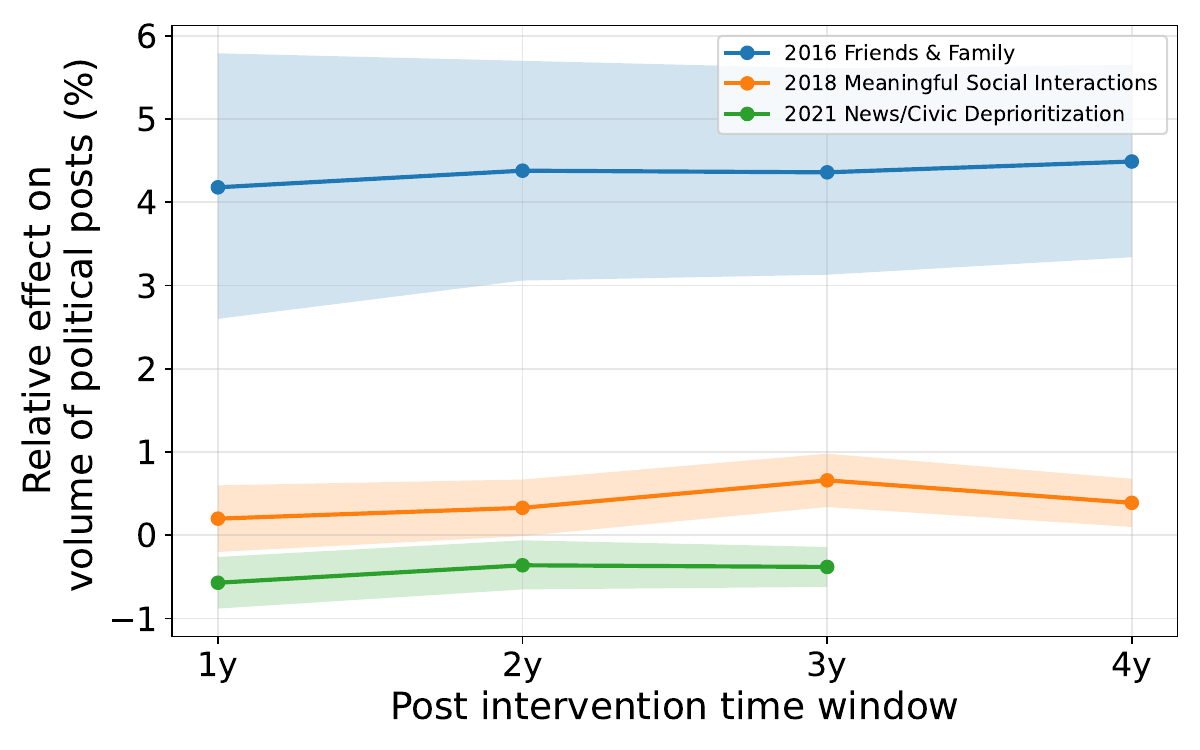}
    \caption{Relative effect of the three interventions on the \textbf{volume} of political posts.}
\label{fig:interventions_combined_political_volume}
\vspace{-\baselineskip}
\end{figure}

\textbf{Results}. The results reveal a complex and often non-linear relationship between platform interventions and the prevalence of political content. Figures~\ref{fig:interventions_combined_political_fraction} and~\ref{fig:interventions_combined_political_volume} show the trends for fraction and volume of political content respectively.

(i) 2016 ``Friends and Family'': The June 2016 update, intended to prioritize personal connections over public content, appears to have paradoxically fostered a more political environment over the long term.
As shown in Figure~\ref{fig:interventions_combined_political_fraction}, the fraction of political posts shows no statistically significant change in the first two years following the June 2016 Friends \& Family change, but increases significantly thereafter (up to +9.1\% at 48 months). By contrast, Figure~\ref{fig:interventions_combined_political_volume} shows that the overall volume of political posts rose significantly even in the first year (+4.2\%, 95\% CI [2.6, 5.8]), suggesting that political content production increased immediately after the intervention. In essence, while the platform aimed to elevate personal content, the 2016 change shifted the pages toward generating more political content, both by increasing its volume and its share of total posts.

(ii) 2018 ``Meaningful Social Interactions'': The January 2018 update, which broadly deprioritized content from Pages, had the most dramatic effect on the composition of users' feeds. The effect on the fraction of political posts was immediate, massive, and sustained. The share of political content surged by 15.5\% in the first year and continued to climb, reaching 22.1\% by the fourth year.
The mechanism behind this shift becomes clear when examining the volume of political posts. The absolute volume of political content barely changed. The estimated effects are less than 1\% across all time windows and are not statistically significant until the 3 year mark.
So, the 2018 intervention filtered out a large quantity of low-engagement, non-political content from Pages (though it was not its sole objective). Political content, which is often inherently controversial and discussion-worthy, likely survived the change because it naturally generates the ``meaningful social interactions'' the update was designed to promote. The result was probably unexpected, increasing the amount of politics on user's feeds.

(iii) 2021 ``News/Civic Deprioritization'': The February 2021 intervention was the most explicit attempt to reduce political content. However, the data shows it had a surprisingly modest and narrowly targeted impact.
%
As intended, the intervention led to a small but statistically significant decline in the volume of political posts. Figure~\ref{fig:interventions_combined_political_volume} shows a consistent decrease of around -0.4\% to -0.6\% across all observed windows. This confirms that the platform did succeed in slightly reducing the total amount of political content being surfaced.
Crucially, however, this reduction in volume did not alter the overall content balance. Figure~\ref{fig:interventions_combined_political_fraction} reveals no statistically significant change in the fraction of political posts. 
So, the 2021 deprioritization likely reduced political and non-political ``civic'' content in roughly equal measure. Unlike the 2018 update, it did not create a compositional shift.
Therefore, while it technically achieved its goal of reducing political volume, its impact on the user's perceived exposure to politics was likely negligible, especially when compared to the profound compositional effects of the 2018 change.

We also find interesting differences in effects of these interventions across demographics. Due to space constraints, we present them in the Appendix (Section~\ref{subsec:interventions_across_demographics}).

\section{Discussion}
This study leverages a large-scale, demographically representative dataset of Facebook activity to construct a high-resolution map of potential exposure to political content in the United States over a decade. By moving beyond platform-provided tools and using a novel data donation methodology, we offer three core contributions. First, we provide robust, population-level estimates of who is exposed to political content, revealing durable demographic divides that constitute separate civic realities on the same platform. Second, we quantify the phenomenon of incidental exposure, showing how political discourse permeates non-political spaces. Finally, we analyze the profound and often counterintuitive impact of platform interventions, demonstrating that algorithmic architecture can be a more powerful driver of politicization than explicit content policy.


Our findings challenge the common narrative that Facebook is an overwhelmingly political environment. Across hundreds of millions of posts, political content accounts for less than one-fifth of the material users could encounter. The majority of potential exposure remains apolitical. This reinforces a growing body of work arguing that the perception of social media as hyper-politicized stems less from overall content supply and more from how salient and emotionally charged political posts are when they do appear~\cite{barbera2020social}. At the same time, we show that political discussion is not confined to explicitly political spaces. Roughly one-third of all political posts originate from non-political pages, confirming that incidental political exposure is a core feature of Facebook’s information environment rather than a fringe phenomenon.

A central finding of our work is that exposure to political content is far from uniform. Instead, it is deeply stratified by age, gender, and ethnicity, creating fundamentally different online experiences. Our ten-year analysis reveals that older users and men are potentially exposed to significantly more political content than younger users and women, while Hispanic users engage with it far less than other groups. These are not fleeting trends but durable, structural features of the Facebook ecosystem, remaining stable across a decade of immense political and social change spanning our dataset.

The stark gap between young and old users, for instance, is likely a product of both life-cycle effects—where political interest often correlates with age—and cohort effects, reflecting fundamentally different media exposure habits and modes of platform engagement among generations \cite{dellaposta2015liberals}. Rather than a single ``digital town square,'' platforms like Facebook host multiple, parallel public spheres with vastly different levels of political saturation. Understanding these baseline differences is a critical prerequisite for any theory of online polarization or misinformation; we cannot speak of a shared ``echo chamber'' when many users are barely in the political conversation to begin with.

Our causal analyses reveal that internal platform interventions can substantially reweight the composition of political exposure, even when they are not explicitly political in intent.
Together, these interventions show that ranking changes reshape \textit{where} political content comes from rather than eliminating it. Even as Facebook reduces exposure from political or news pages, political posts persist in non-political communities. This persistence highlights that incidental exposure is a built-in feature of Facebook’s social architecture, not an artifact of overtly political engagement.

Finally, a major challenge in contemporary computational social science is the reliance on platform-owned APIs and data, which can be restricted. Our approach provides a model for constructing high-quality, representative datasets that empower independent analysis.
The generalizability of this model is considerable and can be adapted for studying content ecosystems on other major platforms like YouTube, or TikTok, where robust APIs may exist but often lack the metadata needed for representative sampling.

\textbf{Future work}. 
Our study identified durable divides in political content exposure across age, gender, and ethnicity. The critical next step is to move beyond these main effects and investigate intersectional dynamics. For instance, how does the online civic reality of a young, college-educated woman differ from that of an older, non-college-educated man? By analyzing the interaction effects between demographics, we can build a far more granular model of political identity and media consumption online. Furthermore, the longitudinal nature of our data collection allows us to track these specific intersectional groups over time. This enables analyses like: Do major events, such as the 2024 election or a social movement, cause these demographic gaps to widen or narrow?

A key finding was the prevalence of political content within non-political pages. Incidental exposure remains one of the least understood but most important aspects of today’s online information environment. Future work should aim to map where and how such exposure occurs across platforms and communities.
This could involve identifying the specific genres of non-political pages (e.g., meme pages, or hobbyist groups, local commerce pages, celebrity fan clubs) that serve as the primary vectors for political content. 
This would help clarify the mechanisms by which users who do not seek out politics are nevertheless drawn into civic discourse.

Our analysis of platform interventions was necessarily retrospective. The next frontier is prospective and experimental. By leveraging our rich demographic and behavioral data, we can develop ``artificial silicon samples''—agent-based models or ``bots'' that accurately simulate the behavior of diverse user populations~\cite{argyle2023out}. These digital twins could be used to create an in silico experimental sandbox to test counterfactuals such as how the impact of the 2018 ``Meaningful Social Interactions'' change would have differed if the user base were younger or more ethnically diverse, or even simulate future interventions like the impact of a policy on political exposure across different demographic groups~\cite{hosseinmardi2024causally}.

\textbf{Limitations}. 
The most significant limitation of our study is that we measure potential exposure (the universe of content produced by pages a user follows) rather than realized exposure (the specific content that algorithmically appears in their News Feed). A user’s feed is a complex and personalized mix of content from multiple sources: public Pages, private Groups, posts from friends and family, and, increasingly, algorithmically recommended content from sources the user does not follow. Our methodology does not capture these other crucial streams of information. Therefore, while our data accurately maps the public-facing information environment a user opts into, it cannot fully represent the personalized ``diet'' of content the platform ultimately serves them. This gap is a fundamental challenge for all external platform research. However, as we mention in Section~\ref{sec:data_collection}, analysis of Facebook's own data (see Figure~\ref{fig:facebook_frac_pages_groups}) indicates a significant chunk of user's feeds containing public Page and Group content, indicating directional validity of our results.

Our analysis is limited to textual and metadata information from public Facebook posts. We do not systematically analyze images, videos, or links, even though they can carry strong political cues or emotional framing. This may cause us to understate the extent and style of political communication, particularly in image- or video-heavy spaces. That said, Facebook remains predominantly text-driven---85\% of public posts contain some textual component---so our estimates capture most of the political discourse likely to reach users. Future work could integrate multimodal analysis to reveal how visual narratives contribute to incidental political exposure.

Finally, care should be taken in interpreting our causal claims. Our BSTS framework provides credible counterfactuals based on pre-intervention trends, but it relies on key assumptions: that no unobserved shocks uniquely affect the post-intervention period, and that covariates remain exogenous. In practice, Facebook’s interventions roll out gradually, often alongside other policy or design changes, and may coincide with external events such as elections or the COVID-19 pandemic. Although we mitigate this with diagnostic checks, placebo tests, and multiple post-period windows, residual confounding remains possible. The temporal overlap among the three interventions also complicates clean attribution, as long-lived effects from one update can spill over into the next. Hence, our results should be interpreted as robust directional evidence of how interventions reshape political exposure, rather than as precise causal estimates of isolated changes.

\bibliographystyle{ACM-Reference-Format}
\bibliography{sample-base}

@misc{pewresearchAmericansSocial,
	author = {Jeffrey Gottfried},
	title = {{A}mericans’ {S}ocial {M}edia {U}se --- pewresearch.org},
	howpublished = {\url{https://www.pewresearch.org/internet/2024/01/31/americans-social-media-use/}},
	year = {2023},
	note = {[Accessed 07 May 2024]},
}

@article{freelon2018computational,
  title={Computational research in the post-API age},
  author={Freelon, Deen},
  journal={Political Communication},
  volume={35},
  number={4},
  pages={665--668},
  year={2018},
  publisher={Taylor \& Francis}
}

@article{keusch2023you,
  title={Do you have two minutes to talk about your data? Willingness to participate and nonparticipation bias in Facebook data donation},
  author={Keusch, Florian and Pankowska, Paulina K and Cernat, Alexandru and Bach, Ruben L},
  journal={Field Methods},
  year={2023},
  publisher={Sage Publications Sage CA: Los Angeles, CA}
}

@article{fraxanet2025analyzing,
  title={Analyzing news engagement on Facebook: tracking ideological segregation and news quality in the Facebook URL dataset},
  author={Fraxanet, Emma and Kaltenbrunner, Andreas and Germano, Fabrizio and G{\'o}mez, Vicen{\c{c}}},
  journal={EPJ Data Science},
  volume={14},
  number={1},
  pages={73},
  year={2025},
  publisher={Springer}
}

@misc{nytimesFacebookDials2021,
	author = {Kevin Roose, Mike Isaac},
	title = {{F}acebook {D}ials {D}own the {P}olitics for {U}sers ({P}ublished 2021) --- nytimes.com},
	howpublished = {\url{https://www.nytimes.com/2021/02/10/technology/facebook-reduces-politics-feeds.html}},
	year = {2021},
	note = {[Accessed 06-10-2025]},
}

@article{brodersen2015inferring,
  title={Inferring causal impact using Bayesian structural time-series models},
  author={Brodersen, Kay H and Gallusser, Fabian and Koehler, Jim and Remy, Nicolas and Scott, Steven L},
  journal={Annals of Applied Statistics},
  year={2015}
}

@article{barbera2020social,
  title={Social media, echo chambers, and political polarization},
  author={Barber{\'a}, Pablo},
  journal={Social media and democracy: The state of the field, prospects for reform},
  pages={34--55},
  year={2020},
  publisher={SSRC Anxieties of Democracy, Cambridge University Press, Cambridge, UK}
}

@inproceedings{garimella2018political,
  title={Political discourse on social media: Echo chambers, gatekeepers, and the price of bipartisanship},
  author={Garimella, Kiran and De Francisci Morales, Gianmarco and Gionis, Aristides and Mathioudakis, Michael},
  booktitle={Proceedings of the 2018 world wide web conference},
  pages={913--922},
  year={2018}
}

@inproceedings{10.1145/3720533.3756893,
author = {Rubinstein, Yair},
title = {Meta Content Library as a Research Tool},
year = {2025},
isbn = {9798400715334},
publisher = {Association for Computing Machinery},
address = {New York, NY, USA},
url = {https://doi.org/10.1145/3720533.3756893},
doi = {10.1145/3720533.3756893},
booktitle = {Adjunct Proceedings of the 36th ACM Conference on Hypertext and Social Media},
pages = {54},
numpages = {1},
location = {
},
series = {HT Adjunct '25}
}

@article{eady2023exposure,
  title={Exposure to the Russian Internet Research Agency foreign influence campaign on Twitter in the 2016 US election and its relationship to attitudes and voting behavior},
  author={Eady, Gregory and others},
  journal={Nature Communications},
  year={2023},
  publisher={Nature Publishing Group UK London}
}

@misc{pew2022twitterpolitics,
  title = {Politics on Twitter: One-Third of Tweets From U.S. Adults Are Political},
  author = {{Pew Research Center}},
  year = {2022},
  institution = {Pew Research Center},
  url = {https://www.pewresearch.org/politics/2022/06/16/politics-on-twitter-one-third-of-tweets-from-u-s-adults-are-political/}
}

@article{costello2016views,
  title={Who views online extremism? Individual attributes leading to exposure},
  author={Costello, Matthew and Hawdon, James and Ratliff, Thomas and Grantham, Tyler},
  journal={Computers in human behavior},
  volume={63},
  pages={311--320},
  year={2016},
  publisher={Elsevier}
}

@article{hobolt2024polarizing,
  title={The polarizing effect of partisan echo chambers},
  author={Hobolt, Sara B and Lawall, Katharina and Tilley, James},
  journal={American Political Science Review},
  volume={118},
  number={3},
  pages={1464--1479},
  year={2024},
  publisher={Cambridge University Press}
}

@article{sultan2024susceptibility,
  title={Susceptibility to online misinformation: A systematic meta-analysis of demographic and psychological factors},
  author={Sultan, Mubashir and Tump, Alan N and Ehmann, Nina and Lorenz-Spreen, Philipp and Hertwig, Ralph and Gollwitzer, Anton and Kurvers, Ralf HJM},
  journal={Proceedings of the National Academy of Sciences},
  volume={121},
  number={47},
  pages={e2409329121},
  year={2024},
  publisher={National Academy of Sciences}
}

@article{guess2019less,
  title={Less than you think: Prevalence and predictors of fake news dissemination on Facebook},
  author={Guess, Andrew and Nagler, Jonathan and Tucker, Joshua},
  journal={Science advances},
  volume={5},
  number={1},
  pages={eaau4586},
  year={2019},
  publisher={American Association for the Advancement of Science}
}

@article{bakshy2015exposure,
  title={Exposure to ideologically diverse news and opinion on Facebook},
  author={Bakshy, Eytan and Messing, Solomon and Adamic, Lada A},
  journal={Science},
  volume={348},
  number={6239},
  pages={1130--1132},
  year={2015},
  publisher={American Association for the Advancement of Science}
}

@article{weeks2017incidental,
  title={Incidental exposure, selective exposure, and political information sharing: Integrating online exposure patterns and expression on social media},
  author={Weeks, Brian E and Lane, Daniel S and Kim, Dam Hee and Lee, Slgi S and Kwak, Nojin},
  journal={Journal of computer-mediated communication},
  volume={22},
  number={6},
  pages={363--379},
  year={2017},
  publisher={Oxford University Press Oxford, UK}
}

@article{flaxman2016filter,
  title={Filter bubbles, echo chambers, and online news consumption},
  author={Flaxman, Seth and Goel, Sharad and Rao, Justin M},
  journal={Public opinion quarterly},
  volume={80},
  number={S1},
  pages={298--320},
  year={2016},
  publisher={Oxford University Press US}
}

@article{guess2023social,
  title={How do social media feed algorithms affect attitudes and behavior in an election campaign?},
  author={Guess, Andrew M and Malhotra, Neil and Pan, Jennifer and Barber{\'a}, Pablo and Allcott, Hunt and Brown, Taylor and Crespo-Tenorio, Adriana and Dimmery, Drew and Freelon, Deen and Gentzkow, Matthew and others},
  journal={Science},
  volume={381},
  number={6656},
  pages={398--404},
  year={2023},
  publisher={American Association for the Advancement of Science}
}

@article{talaga2025changes,
  title={Changes to the Facebook Algorithm Decreased News Visibility Between 2021-2024},
  author={Talaga, Szymon and Wertz, Erin and Batorski, Dominik and Wojcieszak, Magdalena},
  journal={arXiv preprint arXiv:2507.19373},
  year={2025}
}

@article{lazer2020covid,
  title={The COVID States Project: A 50-state COVID-19 survey report\# 26: Trajectory of COVID-19-related behaviors},
  author={Lazer, David and Santillana, Mauricio and Perlis, Roy H and Quintana, Alexi and Ognyanova, Katherine and Green, Jonathan and Baum, Matthew A and Simonson, Matthew and Uslu, Ata A and Chwe, Hanyu and others},
  journal={COVID States Project},
  year={2020}
}

@article{allen2020evaluating,
  title={Evaluating the fake news problem at the scale of the information ecosystem},
  author={Allen, Jennifer and Howland, Baird and Mobius, Markus and Rothschild, David and Watts, Duncan J},
  journal={Science advances},
  year={2020},
  publisher={American Association for the Advancement of Science}
}

@article{hosseinmardi2024causally,
  title={Causally estimating the effect of YouTube’s recommender system using counterfactual bots},
  author={Hosseinmardi, Homa and Ghasemian, Amir and Rivera-Lanas, Miguel and Horta Ribeiro, Manoel and West, Robert and Watts, Duncan J},
  journal={PNAS},
  year={2024},
  publisher={National Academy of Sciences}
}

@article{ohme2023digital,
  title={Digital trace data collection for social media effects research: APIs, data donation, and (screen) tracking},
  author={Ohme, Jakob and Araujo, Theo and Boeschoten, Laura and Freelon, Deen and Reeves, Byron B and Robinson, Thomas N},
  journal={Communication Methods and Measures},
  year={2023},
  publisher={Taylor \& Francis}
}

@misc{crowdtangleAboutCrowdTangle,
	author = {Tess},
	title = {{A}bout {U}s | {C}rowd{T}angle {H}elp {C}enter --- help.crowdtangle.com},
	howpublished = {\url{https://help.crowdtangle.com/en/articles/4201940-about-us}},
	year = {2018},
	note = {[Accessed 10 May 2024]},
}

@article{dellaposta2015liberals,
  title={Why do liberals drink lattes?},
  author={DellaPosta, Daniel and Shi, Yongren and Macy, Michael},
  journal={American Journal of Sociology},
  volume={120},
  number={5},
  pages={1473--1511},
  year={2015},
  publisher={University of Chicago Press Chicago, IL}
}

@misc{MetaTransparencyCenter2024,
  title = {Meta Transparency Center},
  author = {Meta Platforms, Inc.},
  year = {2024},
  howpublished = {\url{https://transparency.meta.com}},
  note = {Accessed: 2025-09-15}
}

@misc{facebook2018,
  author = {Adam Mosseri and Facebook, Inc.},
  title = {News Feed FYI: Bringing People Closer Together},
  year = {2018},
  howpublished = {\url{https://about.fb.com/news/2018/01/news-feed-fyi-bringing-people-closer-together/}},
  note = {Facebook's announcement of their 2018 News Feed algorithm change prioritizing meaningful social interactions (MSI).}
}

@article{sarig2023balance,
  title={balance--a Python package for balancing biased data samples},
  author={Sarig, Tal and Galili, Tal and Eilat, Roee},
  journal={arXiv preprint arXiv:2307.06024},
  year={2023}
}

@article{heawood2018pseudo,
  title={Pseudo-public political speech: Democratic implications of the Cambridge Analytica scandal},
  author={Heawood, Jonathan},
  journal={Information polity},
  volume={23},
  number={4},
  pages={429--434},
  year={2018},
  publisher={IOS Press}
}

@article{burroughs2014facebook,
  title={Facebook and FarmVille: A digital ritual analysis of social gaming},
  author={Burroughs, Benjamin},
  journal={Games and Culture},
  year={2014},
  publisher={Sage Publications Sage CA: Los Angeles, CA}
}

@misc{crowdtangleWhatData,
	author = {Tess},
	title = {{W}hat data is {C}rowd{T}angle tracking? | {C}rowd{T}angle {H}elp {C}enter --- help.crowdtangle.com},
	howpublished = {\url{https://help.crowdtangle.com/en/articles/1140930-what-data-is-crowdtangle-tracking}},
	year = {2024},
	note = {[Accessed 15 May 2024]},
}

@article{argyle2023out,
  title={Out of one, many: Using language models to simulate human samples},
  author={Argyle, Lisa P and Busby, Ethan C and Gubler, Joshua R and Rytting, Christopher and Wingate, David},
  journal={Political Analysis},
  year={2023},
  publisher={Cambridge University Press}
}

@inproceedings{fulay-etal-2024-relationship,
    title = "On the Relationship between Truth and Political Bias in Language Models",
    author = "Fulay, Suyash  and
      Brannon, William  and
      Mohanty, Shrestha  and
      Overney, Cassandra  and
      Poole-Dayan, Elinor  and
      Roy, Deb  and
      Kabbara, Jad",
    editor = "Al-Onaizan, Yaser  and
      Bansal, Mohit  and
      Chen, Yun-Nung",
    booktitle = "Proceedings of the 2024 Conference on Empirical Methods in Natural Language Processing",
    month = nov,
    year = "2024",
    address = "Miami, Florida, USA",
    publisher = "Association for Computational Linguistics",
    url = "https://aclanthology.org/2024.emnlp-main.508/",
    doi = "10.18653/v1/2024.emnlp-main.508",
    pages = "9004--9018"
}

@inproceedings{ding-etal-2023-gpt,
    title = "Is {GPT}-3 a Good Data Annotator?",
    author = "Ding, Bosheng  and
      Qin, Chengwei  and
      Liu, Linlin  and
      Chia, Yew Ken  and
      Li, Boyang  and
      Joty, Shafiq  and
      Bing, Lidong",
    editor = "Rogers, Anna  and
      Boyd-Graber, Jordan  and
      Okazaki, Naoaki",
    booktitle = "Proceedings of the 61st Annual Meeting of the Association for Computational Linguistics (Volume 1: Long Papers)",
    month = jul,
    year = "2023",
    address = "Toronto, Canada",
    publisher = "Association for Computational Linguistics",
    url = "https://aclanthology.org/2023.acl-long.626/",
    doi = "10.18653/v1/2023.acl-long.626",
    pages = "11173--11195"
}

@book{pew2024changing,
  title={Changing partisan coalitions in a politically divided nation},
  author={Pew Research Center},
  year={2024},
  publisher={Pew Research Center}
}

@misc{openai2025gpt5,
  author       = {OpenAI},
  title        = {Introducing GPT-5},
  year         = {2025},
  howpublished = {\url{https://openai.com/index/introducing-gpt-5/}}
}

@misc{meta_llama330_70BInstruct_2024,
  author       = {Meta},
  title        = {Llama-3.3-70B-Instruct},
  year         = {2024},
  howpublished = {\url{https://huggingface.co/meta-llama/Llama-3.3-70B-Instruct}},
  note         = {Release date: December 6, 2024}
}

@misc{pew_harris_trump_supporters_2024,
  author       = {Pew Research Center},
  title        = {The political values of Harris and Trump supporters},
  year         = {2024},
  howpublished = {\url{https://www.pewresearch.org/politics/2024/08/26/the-political-values-of-harris-and-trump-supporters/}},
  note         = {August 26, 2024}
}

@misc{FB_friendsfamily_2016,
  author       = {Mosseri, Adam},
  title        = {Building a Better News Feed for You},
  howpublished = {\url{https://about.fb.com/news/2016/06/building-a-better-news-feed-for-you/}},
  year         = {2016},
  month        = jun # "~29",
  note         = {Facebook Newsroom},
}

\appendix

\section{Appendix}

\subsection{Proportion of content from groups and pages on Facebook feeds}

In the Meta transparency reports~\cite{MetaTransparencyCenter2024} released every quarter (the reports only started in 2021 Q2), they report the fraction of content that is shown on people's Facebook feeds from various sources like friends, pages, groups, etc. These are actually content shown on a user's feeds, not just production or potential exposure content.
As we can see from Figure~\ref{fig:facebook_frac_pages_groups}, prior to 2023, the share of content from groups and pages shown on a user's feed in 2021-2022 (and likely prior to that which is a majority of our data) is consistently around 30-35\%.

\subsection{Content exposure over time}

\begin{figure*}[ht]
    \includegraphics[width=1\linewidth]{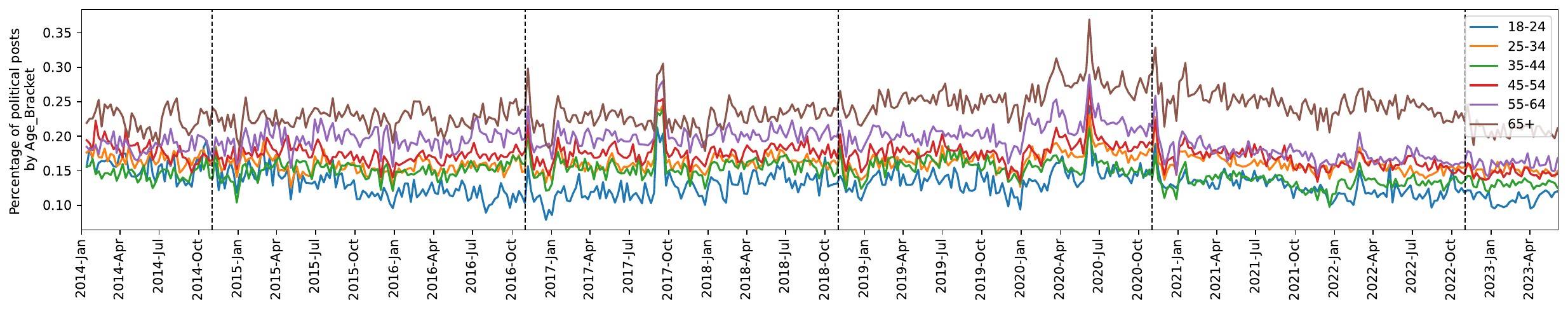}
    \caption{Prevalence of political content per age bracket over 10 years. The black dotted vertical lines are in November of (2014, 2016, 2018, 2020 and 2022) indicating elections/midterms.}
    \label{fig:10y_political_age}
\end{figure*}

\begin{figure*}[ht]
    \includegraphics[width=1\linewidth]{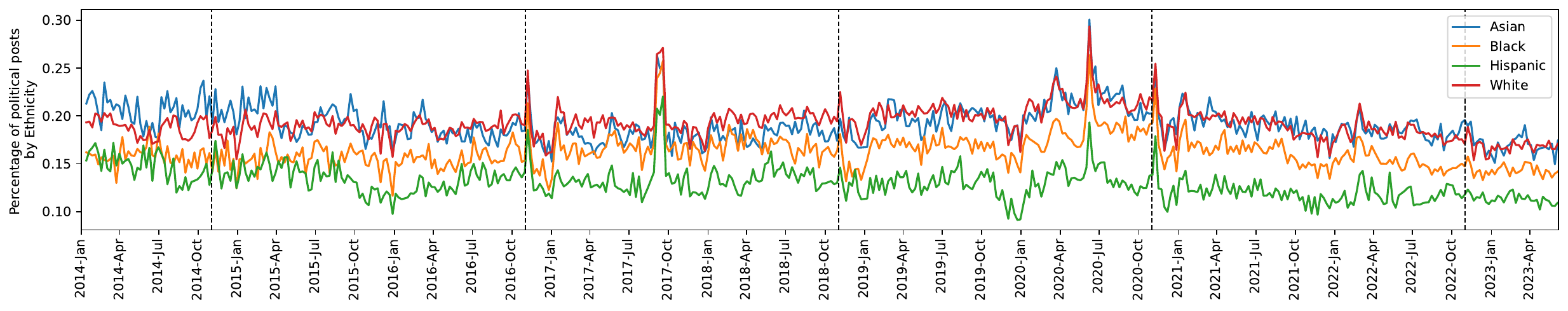}
    \caption{Prevalence of political content per ethnicity over 10 years}
    \label{fig:10y_political_eth}
\end{figure*}

Figures~\ref{fig:10y_political_age},~\ref{fig:10y_political_eth} show the trends in fraction of political content exposure over roughly 10 years span of our dataset. It is interesting that the trends in types of content potentially exposed going back in time remain consistent.

\subsection{Partisan content exposure}

\begin{figure}[ht]
\includegraphics[width=0.7\linewidth, alt={Bar chart showing share of left and right leaning content by ethnicity}]
{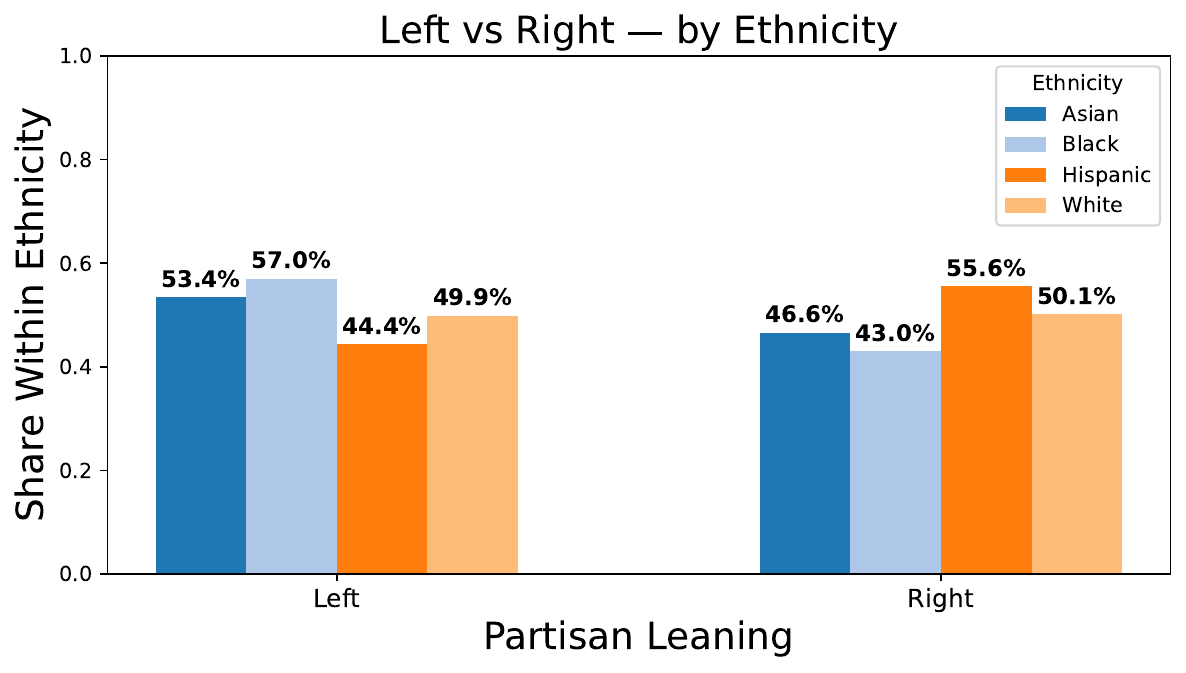}
    \caption{Share of left and right leaning content by ethnicity}
    \label{fig:left_right_overall_ethnicity}
\vspace{-\baselineskip}
\end{figure}

\begin{figure}[ht]
    \centering
    \includegraphics[width=0.7\linewidth]{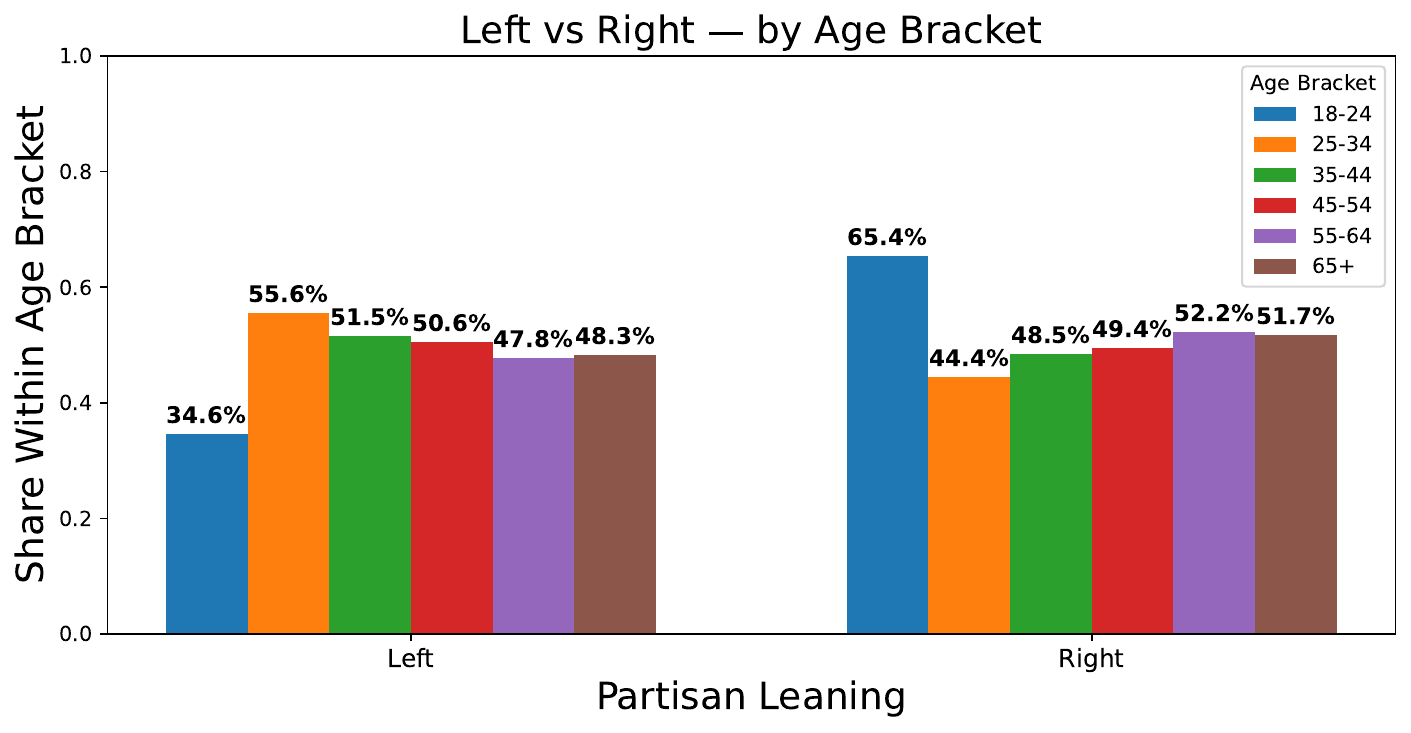}
    \caption{Share of left and right leaning content by age}
    \label{fig:left_right_overall_age}
\vspace{-\baselineskip}
\end{figure}

\begin{figure}[ht]
    \centering
    \includegraphics[width=0.7\linewidth]{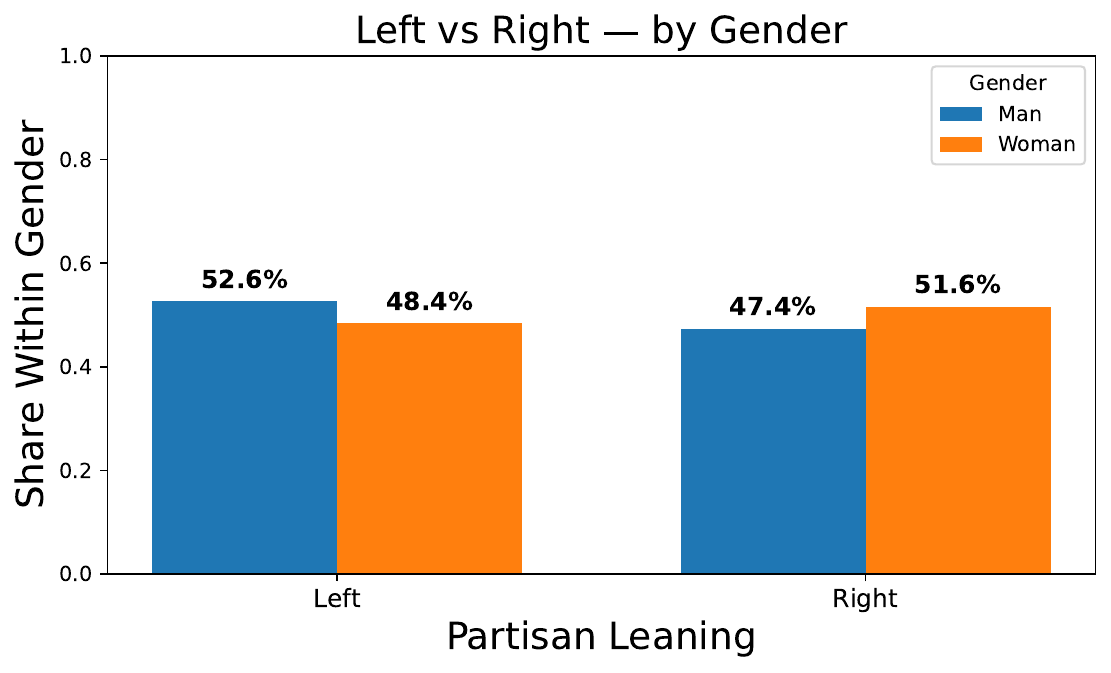}
    \caption{Share of left and right leaning content by gender}
    \label{fig:left_right_overall_gender}
\vspace{-\baselineskip}
\end{figure}

\begin{figure}[ht]
    \centering
    \includegraphics[width=0.7\linewidth]{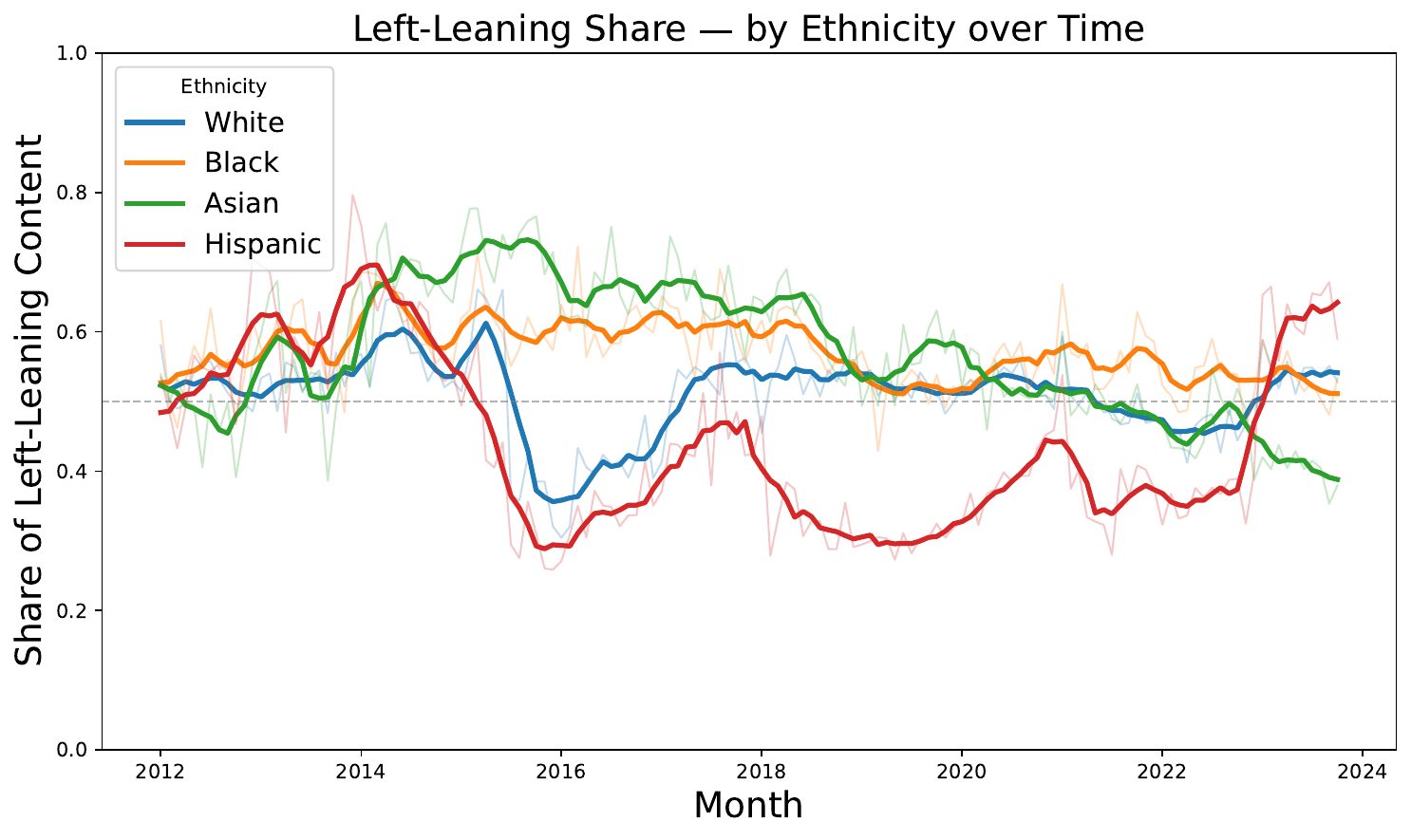}
    \caption{Share of left leaning content by ethnicity over time}
    \label{fig:left_right_overall_ethnicity_time}
\end{figure}

\begin{figure}[ht]
    \centering
    \includegraphics[width=0.7\linewidth]{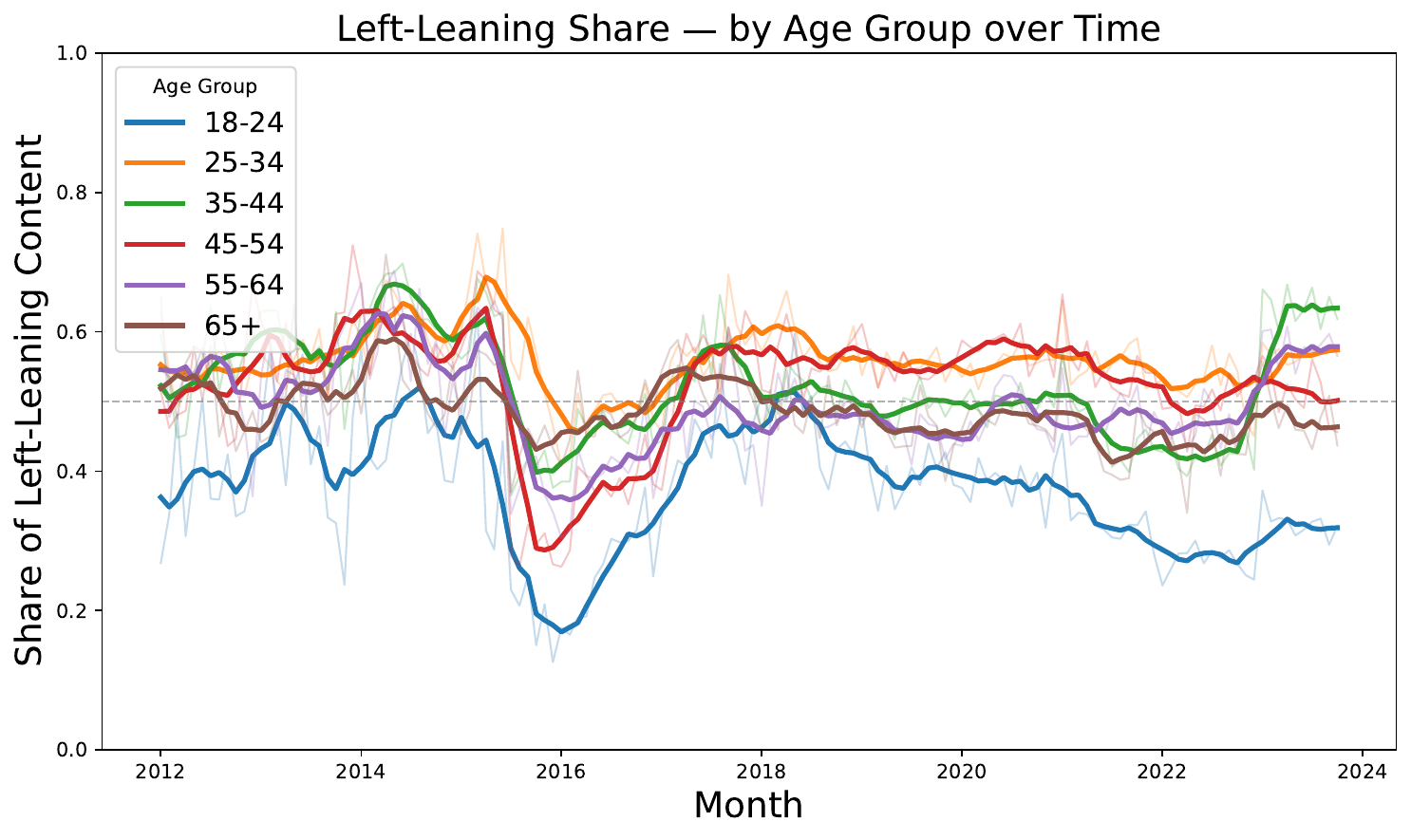}
    \caption{Share of left leaning content by age over time}
    \label{fig:left_right_overall_age_time}
\end{figure}

\begin{figure}[ht]
    \centering
    \includegraphics[width=0.7\linewidth]{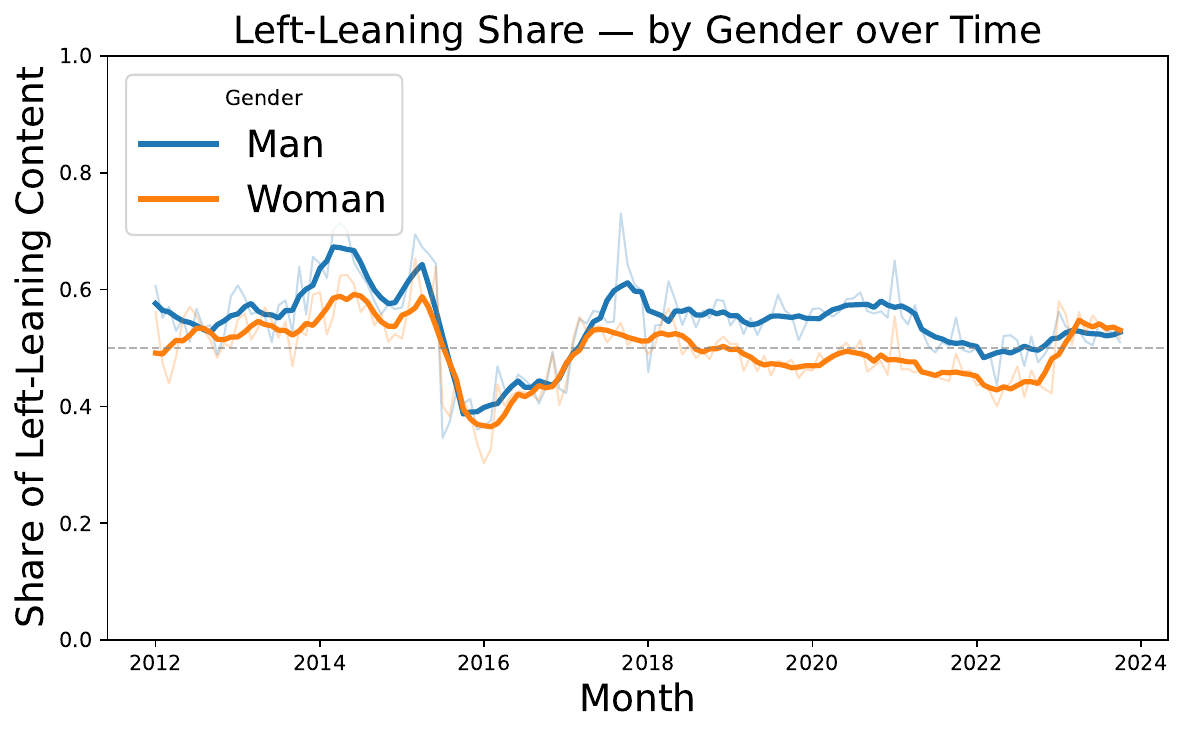}
    \caption{Share of left leaning content by gender over time}
    \label{fig:left_right_overall_gender_time}
\end{figure}

Figures~\ref{fig:left_right_overall_ethnicity},~\ref{fig:left_right_overall_age},~\ref{fig:left_right_overall_gender} show the percentage of potential exposure to political content by each demographic from left or right. Left or right leaning content is defined by a cut off on the leaning score predicted in Section~\ref{sec:llama} with a $<$40 score as left leaning and $>$60 score as right leaning. Gender-wise, men lean slightly more left than women, while women show a slight preference for right-leaning content, though the differences are modest and nearly balanced. Ethnicity reveals more pronounced divergence: Black and Asian users exhibit stronger left-leaning tendencies, whereas Hispanic and White users are more evenly split, with a slight rightward tilt among Hispanics. Age shows the clearest polarization: the youngest group (18–24) leans significantly right (65.4\%) compared to just 34.6\% on the left, while older groups (especially 25–34) lean more left. Figures~\ref{fig:left_right_overall_ethnicity_time},~\ref{fig:left_right_overall_age_time},~\ref{fig:left_right_overall_gender_time} show trends in these partisan exposure over time for more comprehensive understanding. The figures show the share of left leaning content exposure. The inversion across generations suggests generational divides in political alignment, with younger users tilting conservative and middle-aged adults showing more progressive preferences.

\begin{figure}[ht]
    \centering
    \includegraphics[width=0.8\linewidth]{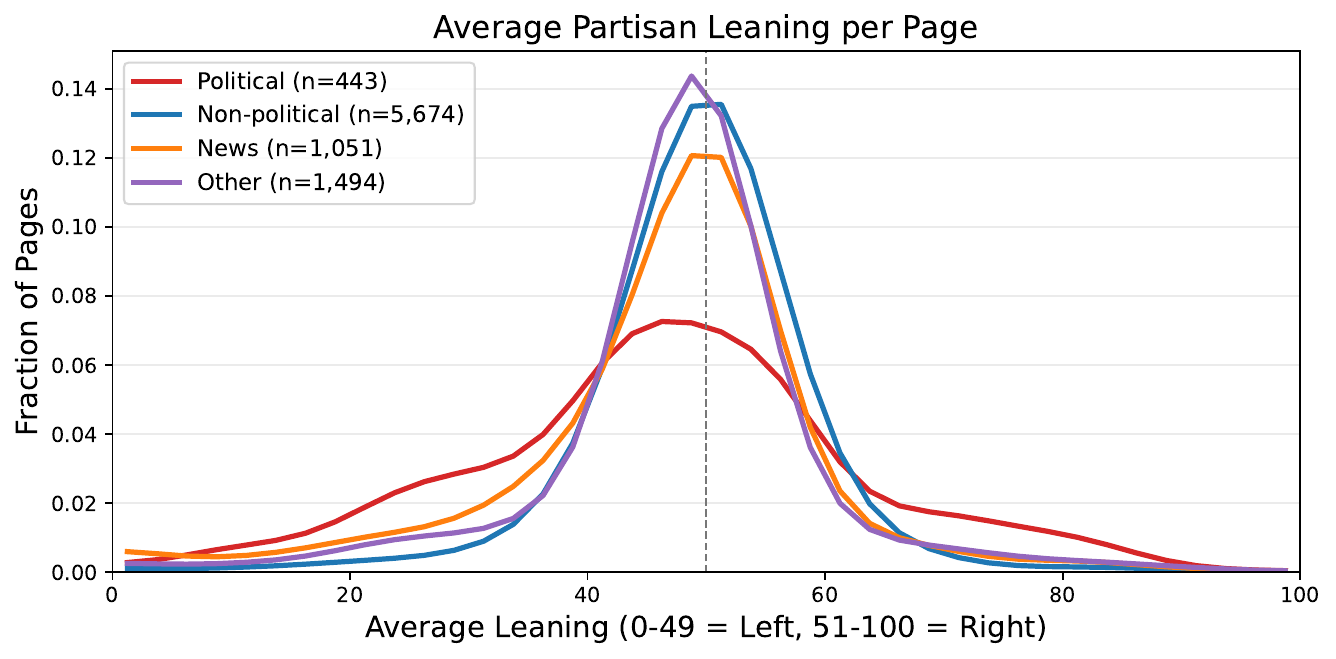}
    \caption{Distribution of average partisan leaning across four page categories.}
    \label{fig:leaning_overlay_fraction}
\vspace{-\baselineskip}
\end{figure}

Figure~\ref{fig:leaning_overlay_fraction} depicts the distribution of average partisan leaning across four page categories—\textit{Political}, \textit{Non-political}, \textit{News}, and \textit{Other} where 0-49 denotes left-leaning and 51-100 right-leaning. Each curve represents the fraction of pages within a given leaning interval, obtained by binning all pages into 40 equal-width bins and dividing by the total number of pages per category. The number of bins for the leaning distribution (40) was selected empirically to provide sufficient granularity across the 0-100 scale while maintaining stable counts per bin across categories. We verified that results were robust to moderate changes (e.g., 30 or 50 bins), indicating that the qualitative shape of each distribution is not sensitive to this choice. To smooth minor fluctuations, a Gaussian kernel was applied while preserving the interpretability of the y-axis as the \emph{fraction of pages}. The plot reveals that \textit{Political} pages are the most ideologically dispersed, spanning both left and right extremes, though with a slightly greater density on the left side of the spectrum. This indicates that while political pages collectively cover a wide ideological range, a larger share of them lean modestly left of center. In contrast, \textit{Non-political} and \textit{News} pages are tightly clustered around the midpoint, suggesting a largely centrist orientation and limited ideological variability. Meanwhile, \textit{Other} pages show a subtle leftward peak and a slightly steeper drop-off on the right, reflecting a mild but noticeable asymmetry in average partisan orientation across this residual category.

\subsection{Demographic Differences Across Platform Interventions}
\label{subsec:interventions_across_demographics}
Across the three major platform interventions, demographic differences in political content exposure and engagement reveal several striking and, at times, counterintuitive patterns. Following the 2016 Friends \& Family update, younger users aged 18--24 experienced a sharp decline in political content exposure ($-19.01\%$ at 12 months), while older users and women saw modest or even positive growth over time---suggesting the intervention may have disproportionately curtailed political exposure among younger users despite similar or increasing volume overall.

The most surprising pattern emerges with the 2018 MSI update, which was explicitly designed to prioritize ``meaningful social interactions'': all demographics saw substantial increases in political content share, with the highest relative effects observed among young adults (18--24: $+40.63\%$), White users ($+26.85\%$), and men ($+24.56\%$). This runs counter to platform narratives that the change would reduce passive news consumption, suggesting instead that political content---perhaps reframed as more ``engaging'' or interpersonal---proliferated even further.

In contrast, the 2021 News Deprioritization efforts show a reversal of this trend, with almost all demographic groups experiencing a decline in political content exposure, particularly older adults (e.g., age 65+: $-13.15\%$ in 2022) and White users ($-9.30\%$). Notably, however, younger users (18--24) are resilient, with no statistically significant decline in political share in 2022 and even a small increase in volume---suggesting either a shift in exposure habits or that the deprioritization filters were less effective for this group.

\subsection{Topics disproportionately exposed to specific demographics}
\label{subsec:topics_disproportionately}

In this section, we identify topics that are disproportionately exposed to specific demographic groups, which showed statistically significant differences compared to others. These differences illuminate the unique preferences and informational needs that characterize diverse demographic segments.
Table~\ref{tab:interest_topics} summarizes results from multiple topics. For brevity, we only included the youngest and oldest age groups. These results indicate statistically significant interest over other demographics.

The findings provide interesting and several surprising insights, while also confirming various stereotypes. For instance, white users were significantly more interested in activities such as art, bird watching, beer brewing and entertainment.
Black users were significantly more interested in sports, family and civil rights. 

It is important to note that our reporting only includes topics where the interest was statistically significant. We abstain from discussing topics where the differences were not substantial, to keep the discussion clear. 
This analysis not only helps in understanding the diverse content preferences across demographic groups but also aids stakeholders in tailoring communication strategies effectively. By understanding these preferences, content creators and policymakers can better address the unique needs of different demographic groups, enhancing engagement and information dissemination.


\begin{table}[ht]
\centering
\caption{Interest Topics by Demographic.}
\label{tab:interest_topics}
\begin{tabular}{l|p{5cm}}
\hline
\textbf{Demographic} & \textbf{Topics of Interest} \\ \hline
White & Art, Bird watching, Beer brewing, British royalty, Film entertainment, Parks, Social issues, US politics, Weather \\ \hline
Black & Basketball, Wrestling, Film entertainment, Kardashians, Civil rights, College sports, Family \\ \hline
Asians & Bollywood, Cricket \\ \hline
Hispanics & Horoscope, British royalty (least) \\ \hline
\hline
Men & Cars, Basketball, Crypto currencies, Gadgets \\ \hline
Women & Bags and accessories, Jewelry, Skin care, Baking, Cooking, Home decor, Family, Horoscope, Kardashians, Animals \\ \hline
\hline
18-24 & Cute babies, Horoscope, Friendship, Climate change \\ \hline
65+ & Migrants on the border, Budlight boycott, Christianity, Lottery, Animals \\ \hline
\end{tabular}
\end{table}

\begin{figure}[ht]
    \centering
    \includegraphics[width=0.7\linewidth]{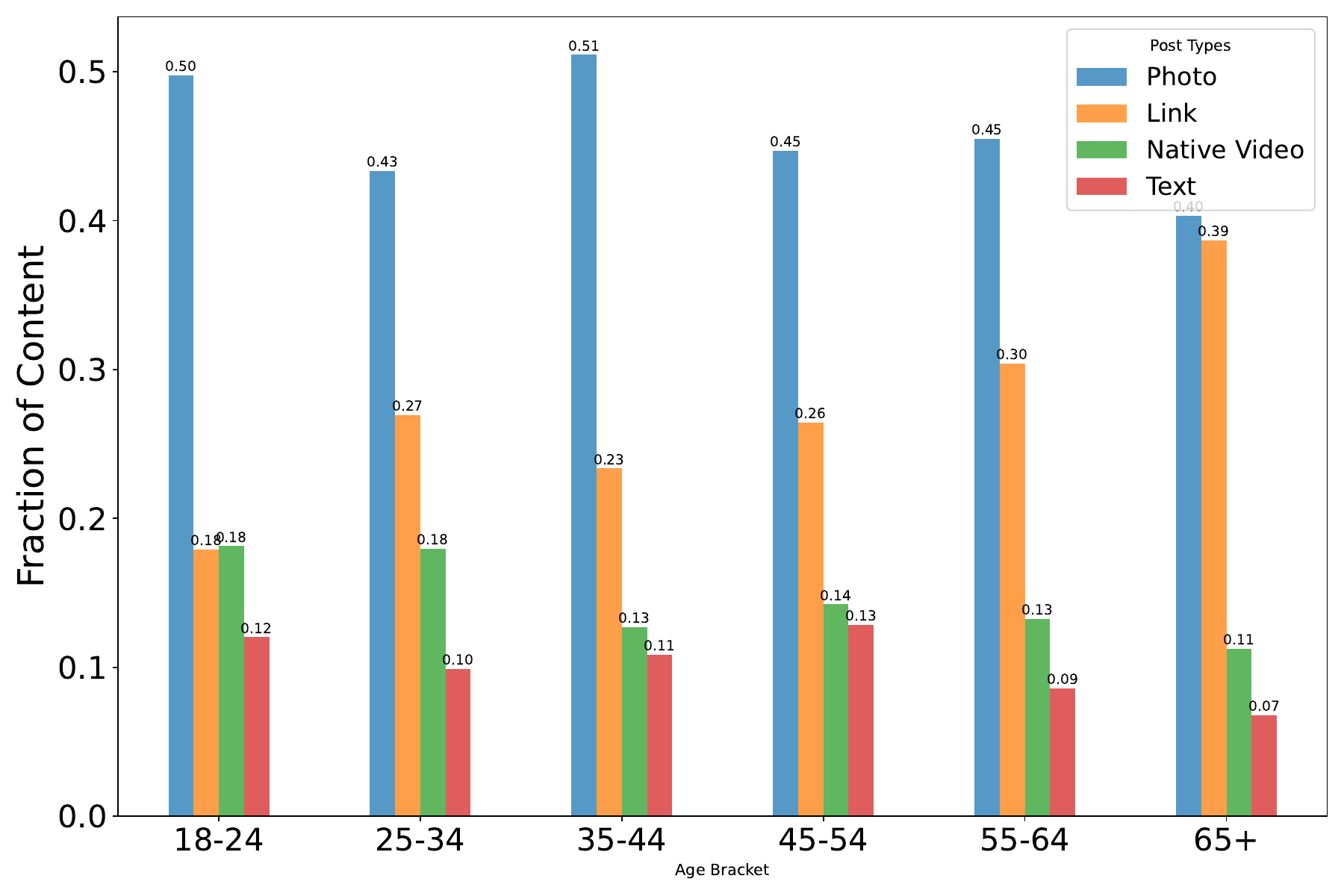}
    \caption{Content types by age group.}
    \label{fig:content_types_age_groups}
\vspace{-\baselineskip}
\end{figure}

\begin{figure}[ht]
    \centering
    \includegraphics[width=0.7\linewidth]{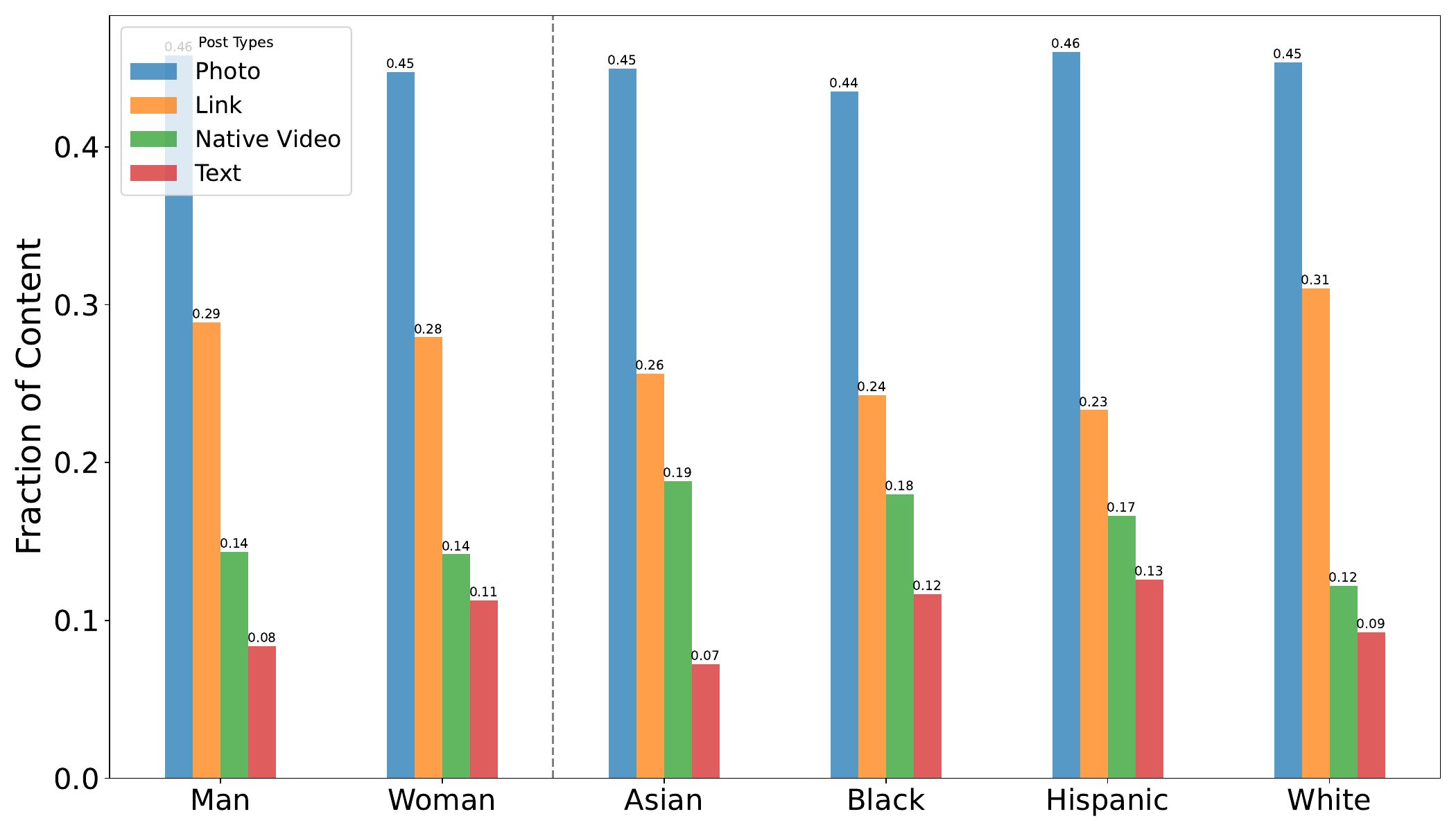}
    \caption{Content types by gender and ethnicity.}
    \label{fig:content_types_gender_ethnicity}
\vspace{-\baselineskip}
\end{figure}

\begin{figure}[ht]
    \centering
    \includegraphics[width=0.7\linewidth]{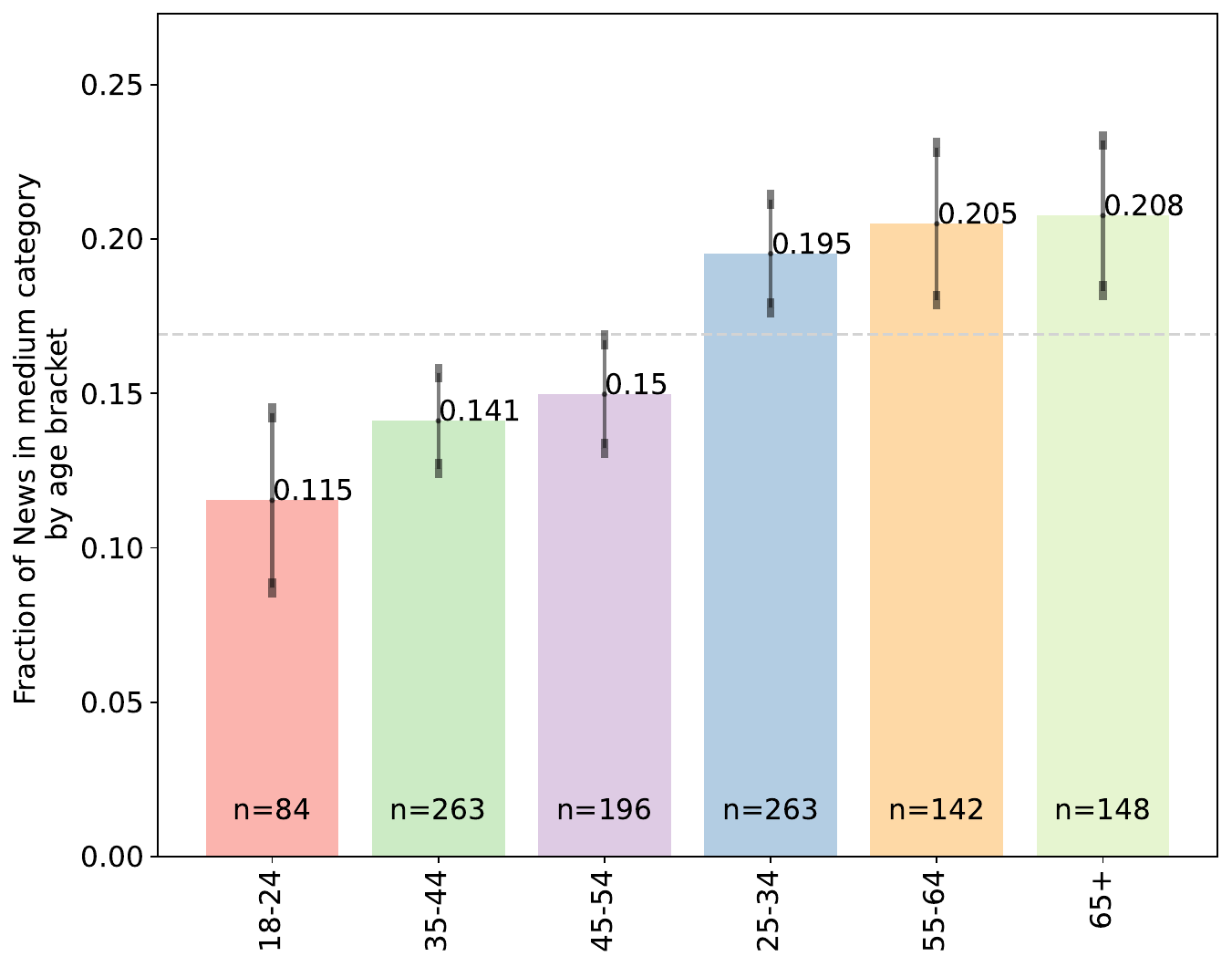}
    \caption{News exposure by age group.}
    \label{fig:news_consumption_age}
\vspace{-\baselineskip}
\end{figure}

\subsection{Heterogeneity in content exposure across groups}
\label{subsec:heterogeneity_content_consumption}

This section digs into the varied modalities through which different demographic groups are exposed to information, highlighting significant differences in content preferences that are not only relevant but also consequential for studies related to information dissemination and misinformation. 

We observed distinct patterns in the way content is exposed to different age groups. As illustrated in Figure~\ref{fig:content_types_age_groups}, older users, particularly those aged 65 and above, exhibit a pronounced preference for link-based content, exposing nearly double the amount of such content compared to the 18-24 age group. In contrast, younger users show a substantial inclination towards video content, reflecting a dynamic shift in engagement as technology and media consumption habits evolve. The heterogeneity extends beyond age and into ethnic differences in content exposure. Figure~\ref{fig:content_types_gender_ethnicity} reveals that White users are less likely to get exposed to video content, accounting for only 12\% of their exposure, compared to 18-19\% for other ethnic groups. Conversely, White users engage more frequently with link-based content, at a rate 5-8\% higher than that observed in other demographics. Intriguingly, our analysis indicates no significant disparities in content exposure patterns across genders.

\end{document}